\documentclass[12pt]{article}
\usepackage{axodraw}
\usepackage{graphicx}

\usepackage{amssymb,amsmath,hhline}

\advance\hoffset by -15mm
\newcommand{\lsim} {\buildrel < \over {_\sim}}
\newcommand{\gsim} {\buildrel > \over {_\sim}}

\begin{document}

\title{Probing Grand Unification Through Neutrino Oscillations, Leptogenesis, and Proton Decay
\thanks{Invited talks presented at the Erice School (September 2002), Neutrino Conference (Stony Brook, October 2002) and PASCOS Conference (Mumbai, January 2003), to appear in the respective Proceedings}
}

\vfill

\author{Jogesh C. Pati\\
Department of Physics, University of Maryland, \\
College Park  MD 20742}

\maketitle
\begin{abstract}
Evidence in favor of supersymmetric grand unification including that based on
the observed family multiplet-structure, gauge coupling unification, neutrino
oscillations, baryogenesis, and certain intriguing features of quark-lepton
masses and mixings is noted. It is argued that attempts to understand (a) the
tiny neutrino masses (especially $\Delta m^2 (\nu_2 -\nu_3)$), (b) the baryon
asymmetry of the universe (which seems to need leptogenesis), and (c) the
observed features of fermion masses such as the ratio $m_b /m_\tau$, the
smallness of $V_{cb}$ and the maximality of $\Theta_{\nu_\mu \nu_\tau}^{osc}$,
seem to select out the route to higher unification based on an effective
string-unified $G(224)=SU(2)_L \times SU(2)_R \times SU(4)^c$ or
$SO(10)$-symmetry, operative in 4D, as opposed to other alternatives.

A predictive framework based on an effective $SO(10)$ or
$G(224)$ symmetry possessing supersymmetry is presented
that successfully describes the masses and mixings of all fermions including
neutrinos. It also accounts for the observed baryon asymmetry of the universe
by utilizing the process of leptogenesis, which is natural to this framework.
It is argued that a conservative upper limit on the proton lifetime within
this $SO(10)/G(224)$-framework, which is so far most successful, is given by
$(\frac{1}{3}-2)\times 10^{34}$ years. This in turn strongly suggests that
an improvement in the current sensitivity by a factor of five to ten
(compared to SuperK) ought to reveal proton decay. Implications of this
prediction for the next-generation nucleon decay and neutrino-detector are
noted.
\end{abstract}
\vfill
\thispagestyle{myheadings}
\markright{UMD-PP-03-048}
\newpage


\begin{section}{Introduction And An Overview}
Since the discoveries (confirmations) of the atmospheric \cite{sk}
and solar neutrino oscillations \cite{sno,bahcall}, the neutrinos have
emerged as being among the most effective probes into the nature of
higher unification.  Although almost the feeblest of all the
entities of nature, simply by virtue of their tiny masses, they
seem to possess a subtle clue to some of the deepest laws of
nature pertaining to the unification-scale as well as the nature
of the unification-symmetry.  In this sense, the neutrinos provide
us with a rare window to view physics at truly short distances. As
we will see, these turn out to be as short as about $10^{-30}$~cm.
Furthermore, it appears most likely that the origin of their tiny
masses may be at the root of the origin of matter-antimatter
asymmetry in the early universe.  In short, the neutrinos may well
be crucial to our own origin!

The main purpose of my talk here today will be to present the intimate links
that exist in the context of supersymmetric grand unification between the 
following phenomena: 
(i) neutrino oscillations, (ii) the masses and mixing of quarks and
charged leptons, (iii) gauge coupling unification, (iv) baryogenesis
via leptogenesis, and last but not least (v) proton decay.

To set the background for a discussion along these lines, let us
first recall that with only left-handed neutrinos, the standard model
based on the gauge symmetry
\(SU(2)_{L} \times U(1)_{Y} \times SU(3)^{c}\),
despite its numerous successes, fails to account for the magnitude
of the mass-difference square
\(\Delta m^{2}(\nu_{2} - \nu_{3}) \sim (1/20\ \mathrm{eV})^{2}\)
observed at Superkamiokande \cite{sk}.  Incorporating effects of
quantum gravity,%
\footnote{See, e.g., S. Weinberg, \emph{Phys. Rev. Lett.}
\textbf{43}, 1566 (1979); \emph{Proc. XXVI Int'l Conf. on High
Energy Physics}, Dallas, TX, 1992; E. Akhmedov, Z. Berezhiani and
G. Senjanovic, \emph{Phys. Rev.} \textbf{D47}, 3245 (1993).
Assuming that quantum gravity could induce violation of lepton
number, one may allow for an effective non-renormalizable operator
of the form \(\lambda_{L} LLHH/M_{pl} + \mathrm{h.c.}\), scaled by
\(M_{pl} = 1.2 \times 10^{19}\ \mathrm{GeV}\) with \(\left< H
\right> \approx 250\ \mathrm{GeV}\). Such an operator would,
however, yield a rather small Majorana mass \(m(\nu_{L}) \sim
10^{-5}\ \mathrm{eV}\). for the left-handed neutrinos, even for a
maximal \(\lambda_{L} \sim 1\), as mentioned in the text.} the
standard model can lead to a neutrino-mass $\sim 10^{-5}\
\mathrm{eV}$, which is, however, too small to account for the
SuperK effect.  One can in fact argue that, to understand the
magnitude of the SuperK effect in any natural way, one would need
\emph{new physics beyond the standard model} at an effective
mass-scale $\sim 10^{15}\ \mathrm{GeV}$, rather than at the Planck
scale $\sim 10^{19}\ \mathrm{GeV}$ \cite{jcp}. Interestingly
enough, one can link this effective mass-scale to the scale of
meeting of the three gauge couplings (to be discussed below) which
is around $2 \times 10^{16} \mathrm{GeV}$. That, in turn, hints at
a link between the physics of neutrino-oscillations and grand
unification!

The idea of ``grand unification'' was introduced in the early
1970's \cite{JCPandAS, GeorgiGlashow, GQW}, purely on aesthetic
grounds, in order to remove certain conceptual shortcomings of the
standard model.  Over the years, a set of key observations ---
some old and some new --- have come to light, which together
provide strong evidence in favor of this idea. Some of the
observations in fact support the idea of both grand unification
and low-energy supersymmetry \cite{susyWessZumino, WittenKalle}.
The evidence includes:
\begin{enumerate}
\item The observed family multiplet-structure --- in particular
the fact that the five (apparently disconnected) multiplets of the
SM belonging to a family neatly become parts of a whole ---
\emph{a single multiplet} --- under grand unification, with
\emph{all} their quantum numbers predicted precisely as observed.
\item The observed quantization of electric charge and the fact
that the electron and the proton have exactly equal but opposite
charges. \item The dramatic meeting of the three gauge couplings
that is found to occur at a scale $M_{X} \approx 2 \times 10^{16}
\mathrm{GeV}$, when they are extrapolated from their values
measured at LEP to higher energies, in the context of
supersymmetry \cite{susyunif}. \item The tiny neutrino masses of
the sort suggested by the discoveries/confirmations of atmospheric
and solar neutrino oscillations. These, as we will see, not only
go well with the scale of unification $M_{X}$ mentioned above but
also help select out a class of unification-symmetries which
provide the right-handed neutrinos $(\nu^{\prime}_{R} s)$ as a
compelling feature and B-L as a local symmetry. \item Certain
intriguing features of the masses and mixings of the quarks and
leptons, including the relation $m_{b} (M_X) \approx m_{\tau}$ and
the largeness of the $\nu_{\mu} -\nu_{\tau}$ oscillation angle
$(\sin^2 2\theta^{osc}_{\nu_{\mu} \nu_{\tau}} \geq 0.92)$ together
with the smallness of $V_{cb} (\approx 0.04)$ \cite{BPW1}. \item
And last but not least, the likely need for leptogenesis
\cite{Vanagidalepto, KRS} to account for the observed
baryon-asymmetry of the universe, which seems to require once
again the existence of superheavy right-handed neutrinos
$(\nu^{\prime}_{R} s)$ and B-L as a local symmetry.
\end{enumerate}

All of these features including the tiny neutrino masses and the observed
baryon-asymmetry can be understood simply, and even quantitatively, within
the concept of supersymmetric grand unification based on an \emph{effective
symmetry in four dimensions}, that is either 
\begin{displaymath}
G(224) = SU(2)_{L} \times SU(2)_{R}\times SU(4)^{C} 
\mbox{ }\mbox{ }\cite{JCPandAS}
\end{displaymath}
\begin{displaymath}
\mbox{or }\mbox{ }\mbox{ }\mbox{ }\mbox{ }\mbox{ }\mbox{ }
SO(10)\mbox{ }\mbox{ }\cite{SO(10)}.
\end{displaymath}
Believing in a unified theory of all forces including gravity, it
is of course attractive to presume that such an effective symmetry
in 4D ($G(224)$ or $SO(10)$) has its origin from a string theory
\cite{stringth} or the M-theory \cite{Mth}. I will comment in Sec.
2 that, in the context of a string theory with the string-scale
being close to the GUT-scale, the observed coupling unification
may be understood even if the effective symmetry in 4D, below the
string scale, is non-simple like $G(224)$. A
string-derived $G(224)$-solution may, however, have an advantage
over an $SO(10)$-solution in that it can neatly avoid the
so-called doublet-triplet splitting problem (generic to SUSY GUTs,
see Sec. 2). Motivated by the desire to avoid this problem, there
have in fact been several attempts in the literature (many rather
recent) which successfully obtain semi-realistic
$G(224)$-solutions in 4D from compactification of a string theory
\cite{StringG(224)}, or of an effective five or six dimensional
GUT-theory \cite{5DG(224)}. For most purposes, in particular
for considerations of fermion masses, neutrino-oscillations, and
leptogenesis, the symmetries $G(224)$ and $SO(10)$ provide
essentially the same advantages. Differences between them in
considerations of proton decay will be noted in Sec. 5.

Let us first recall the new features (relative to the SM) which are introduced
through the symmetry $G(224)$ \cite{JCPandAS}. Subject to left-right discrete symmetry
($L\leftrightarrow R$), which
is natural to $G(224)$, all members of the electron family become parts of a
single left-right self-conjugate multiplet, consisting of:
\begin{equation}
F^{e}_{L,R} = \left[
\begin{array}{cccc}
{u_{r}} & {u_{y}} & {u_{b}} & {\nu_{e}} \\
{d_{r}} & {d_{y}} & {d_{b}} &{{e}^{-}}
\end{array}
\right]_{L,R}.
\label{e3}
\end{equation}
The multiplets $F^{e}_{L}$ and $F^{e}_{R}$ are left-right
conjugates of each other and transform respectively as (2,1,4) and
(1,2,4) of $G(224)$; likewise for the muon and the tau families. The symmetry
$SU(2)_{L}$ treats each column of $F^{e}_{L}$ as a doublet; likewise
$SU(2)_{R}$ for $F^{e}_{R}$.  The symmetry $SU(4)$-color unifies quarks and
lepotons by treating each row of
$F^{e}_{L}$ \emph{and}
$F^{e}_{R}$ as a quartet; \emph{thus lepton number is treated as
the fourth color}.  As mentioned above, because of the parallelism between
$SU(2)_{L}$ and $SU(2)_{R}$, the symmetry $G(224)$ naturally permits the
notion that the fundamental laws of nature possess a left $\leftrightarrow$
right discrete symmetry (i.e.\ parity invariance) that interchanges
\(F_{L}^{e} \leftrightarrow F_{R}^{e}\) and \(W_{L} \leftrightarrow W_{R}\).
With suitable requirements on the Higgs sector, observed parity violation
can be attributed, in this case, entirely to a spontaneous breaking of
the $L \leftrightarrow R$ discrete symmetry \cite{RNMandJCP}.

Furthermore, the symmetry $G(224)$ introduces an elegant charge
formula:
\(Q_{em} = I_{3L} + I_{3R} + (B-L)/2,\)
that applies to all forms of matter (including quarks and leptons of all six
flavors, Higgs and gauge bosons).  Note that the weak hypercharge of the
standard model, given by \(Y_{W} = I_{3R} + (B-L)/2\), is now completely
determined for all members of a family.  Quite clearly, the charges
$I_{3L}$, $I_{3R}$, and B-L, being generators respectively of 
$SU(2)_{L}$ $SU(2)_{R}$, and $SU(4)^{c}$, are quantized; so also then is
the electric charge $Q_{em}$.  Using the expression for $Q_{em}$, one can
now explain why the electron and the proton have exactly equal but opposite
charges.

Note also that postulating either $SU(4)$-color
or $SU(2)_{R}$ forces one to introduce a right-handed neutrino
($\nu_{R}$) for each family as a singlet of the SM symmetry.
\emph{This requires that there be sixteen two-component fermions in each
family, as opposed to fifteen for the SM}. Furthermore, $SU(4)$-color
possesses B-L as one of its generators. This in turn helps to protect the
Majorana masses of the right-handed neutrinos from being of the order string
or Planck-scale.\footnote{Without a protection by a local symmetry,
$\nu^\prime_R s$ (being singlets of the SM) are likely to acquire Majorana
masses of the order string or Planck scale through effects of quantum
gravity. Such ultraheavy $\nu_R$-masses would, however, lead via the seesaw
mechanism (see later), to too small masses for the light neutrinos 
($\leq 10^{-5} \mathrm{eV}$) and
thereby to too small a value for $\Delta m^2_{23}$ compared to observation.
Hence the need for B-L as an effective local symmetry in 4D near the string
scale.} In addition, $SU(4)$-color provides the Dirac mass of the
tau-neutrino  by relating it to the top-quark mass at the unification-scale,
and simultaneously the mass of the
bottom quark in terms of that of the tau-lepton.  In short, $SU(4)$-color
introduces \emph{three characteristic features} --- i.e.
\begin{enumerate}
\item the right-handed neutrinos as a compelling feature,
\item B-L as a local symmetry, and
\item the two GUT-scale mass relations:
\end{enumerate}
\begin{equation}
\begin{array}{lcr}
m_{b}(M_{X}) \approx m_{\tau} & \mbox{and} &
m(\nu^{\tau}_{\mathrm{Dirac}}) \approx m_{\mathrm{top}}(M_{X}) \label{masses}
\end{array}
\end{equation}

These two relations arise from the SU(4)-color preserving leading 
entries in the fermion
mass-matrices (see Sec. 3) which contribute to the masses of the
third family. The sub-leading corrections to the fermion mass
matrices that arise from $SU(4)$-color-breaking in the
(B-L)-direction turn out to be important for the masses and
mixings of the fermions belonging to the first two families
\cite{BPW1}. As we will see, these three ingredients, as well as
the SUSY unification-scale $M_{X}$, play \emph{crucial roles} in
providing us with an understanding of the tiny masses of the
neutrinos as well as of the baryon-asymmetry of the universe, by
utilizing respectively the seesaw mechanism \cite{seesaw} and the
idea of leptogenesis \cite{Vanagidalepto}. \emph{The success of
the predictions in this regard (see below), speaks in
favor of the seesaw mechanism and suggests that the effective
symmetry in 4D, below the string-scale, should contain
$SU(4)$-color}.

Now the minimal symmetry containing $SU(4)$-color on the one hand and also
possessing a rationale for the quantization of electric charge on the other
hand is
provided by the group $G(224)$.  The group $G(224)$ being
isomorphic to \(SO(4) \times SO(6)\) embeds nicely into the simple
group $SO(10)$. The group $SO(10)$, which historically was proposed 
after the suggestion of G(224), of course retains all the
advantages of $G(224)$, in particular the features (a)-(c) listed
above. The interesting point is that $SO(10)$ even preserves the
16-plet multiplet structure of $G(224)$ by putting $\{F_L +
(F_R)^c \}$ as its spinorial 16-dimensional representation,
thereby avoiding the need for any new matter fermions. By
contrast, if one extends $G(224)$ to the still higher symmetry
$E_{6}$ \cite{E6}, one must extend the family-structure from a 16 to
a 27-plet, by postulating additional fermions.

Now utilizing the three ingredients (1), (2), and (3) listed above
(thus assuming that $SU(4)$-color holds in 4D near the GUT-scale),
together with the SUSY unification-scale ($M_{X}$) and the seesaw
mechanism, one arrives at a set of predictions (see Sec. 3) which
include \cite{BPW1}:
\begin{eqnarray}
m_b (m_{b}) & \approx & 4.7-4.9\ \mathrm{GeV} \nonumber \\
m (\nu_{3}) & \approx & (\frac{1}{24}\ \mathrm{eV})(\frac{1}{2}-2) \nonumber \\
\sin^2 2\theta^{\mathrm{osc}}_{\nu_{\mu} \nu_{\tau}} & \approx & 0.99 \nonumber \\
V_{cb} & \approx & 0.044 \label{predictions}
\end{eqnarray}
Each of these predictions agrees remarkably well with
observations.The most intriguing feature is that this framework
provides a compelling reason for why $V_{cb}$ is so small
($\approx 0.04$), and simultaneously why 
\(\sin^{2} 2\theta_{\nu_{\mu}\nu_{\tau}}\) 
is so large ($\approx 1$), both in accord 
with observations.  It is worth noting that the
last two results, showing a sharp difference between $V_{cb}$ and
$\theta_{\nu_{\mu}\nu_{\tau}}$, go against the often expressed
(naive) view that the quark-lepton unification should lead to
similar mixing angles in the quark and lepton sectors.  Quite to
the contrary, as we will see in Sec. 3, the minimal Higgs system
provides a natural breaking of $SU(4)$-color along the
(B--L)-direction which particularly contributes to a mixing between the
second and the third families \cite{BPW1}.  That in turn provides a compelling
\emph{group-theoretical reason} for a distinction between the
masses and mixings of the quarks and leptons as in fact observed
empirically.

One important consequence of having an effective $G(224)$ or 
$SO(10)$-symmetry in 4D is that spontaneous breaking of such a symmetry
(thereby of B-L) into the SM symmetry naturally generates Majorana masses of
the RH neutrinos that are of order GUT-scale or smaller.  In correlation
with the flavor symmetries which provide the hierarchical masses of the quarks
and the leptons, the Majorana masses of the three RH neutrinos are found
to be \cite{JCPlepto}:
\((M_{N_{1}}, M_{N_{2}}, M_{N_{3}}) \approx
(10^{15}, 2 \times 10^{12}, (1/3-2) \times 10^{10}) \mbox{ GeV}\).
Given lepton number (thus of B-L) violation associated with these 
Majorana masses, and C and CP violating 
phases that generically arise in the Dirac
and/or Majorana mass-matrices, the out of equilibrium decays of the lightest
of these heavy RH neutrinos (produced after inflation\footnote{The
lightest $N_{1}$ may be produced either thermally after reheating or
non-thermally through inflaton-decay during reheating.  Both
possibilities are considered in Sec.~3.}) into $l+H$ and
$\bar{l}+\bar{H}$, and the corresponding SUSY modes, generates a 
lepton-asymmetry.  The latter is then converted into a 
baryon-asymmetry by the electroweak sphaleron process 
\cite{Vanagidalepto,KRS}.  In conjunction with an understanding of the
fermion-masses and neutrino-oscillations (atmospheric and solar),
the baryon excess thus generated is found to be (see Sec.~4 and
Ref.~\cite{JCPlepto}:
\begin{equation}
Y_{B} = \left( \frac{n_{B} - n_{\bar{B}}}{n_{s}} \right) \approx
(\sin 2\phi_{21}) (7-100)\times 10^{-11}
\label{baryonexcess}
\end{equation}
While the relevant phase angle $\phi_{21}$ arising from C and
CP-violating phases in the Dirac and Majorana mass-matrices of the
neutrinos is not predictable within the framework, it is rather
impressive that for plausible and natural values of the phase
angle $\phi_{21} \approx \frac{1}{2} -\frac{1}{20}$ (say), the
calculated baryon excess $Y_B$ agrees with the observed value
based on big bang nucleosynthesis \cite{BBN} and CMB data
\cite{CMB}.  This may be contrasted from many alternative
mechanisms, such as GUT and electroweak baryogenesis, which are
either completely ineffective (owing to inflation and gravitino
contraint) or yield too small a baryon excess even for a maximal
phase. For a recent review and other relevant references on the
topic of baryogenesis, see Ref.~\cite{DineKusenko}.

It should be stressed that the five predictions shown in
Eqs(\ref{predictions}) and (\ref{baryonexcess}), together make a
crucial use of the three features (a)-(c) listed in Eq.
(\ref{masses}), as well as of the SUSY unification-scale $M_{X}$
and the seesaw mechanism. Now the properties (a)-(c) are the
distinguishing features of the symmetry $G(224)$. They are of
course available within any symmetry that contains $G(224)$ as a
subgroup. Thus they are present in $SO(10)$ and $E_6$, though not
in $SU(5)$. Effective symmetries like $[SU(3)]^{3}$ \cite{su33} or
\(SU(2)_L \times SU(2)_{R} \times U(1)_{B-L} \times SU(3)^{c}\)
\cite{2213} possess the first two features (a) and (b) but not
(c). Flipped $SU(5) \times U(1)$ \cite{flip} on the other hand
offers (a) and (b) but not the relation $m_{b} (M_X) \approx
m_{\tau}$, which, however, is favored empirically.

The empirical success of the features (1)-(6), including
specifically the predictions listed in Eqs.~(\ref{predictions}) and
(\ref{baryonexcess}), seems to be non-trivial. Together they make
a strong case for both the \emph{conventional ideas of
supersymmetric grand unification}\footnote{By ``conventional'' I
mean gauge coupling unification occurring in 4D at a scale of few
$\times 10^{16}\ \mathrm{Gev}$ (for the case of MSSM), with the
string-scale being somewhat larger. This is to be contrasted from
the case of large extra dimensions with unification occurring at
the TeV-scale, on the one hand, or from 5D GUTs possessing
unification only in higher dimensions leading to the SM symmetry
$G(213)$ in 4D, on the other hand.} \emph{and simultaneously for
the symmetry $G(224)$ or $SO(10)$ being relevant to nature in
four dimensions, just below the string scale}.

As mentioned before, the main purpose of my talk here will be to
present the intimate links that exist, in the context of
supersymmetric grand unification based on an effective $G(224)$ or
$SO(10)$ symmetry, between (i) neutrino oscillations, (ii) the
masses and mixings of quarks and charged leptons, (iii) gauge
coupling unification, (iv) baryogenesis via leptogenesis, and last
but not least (v) proton decay.

Perhaps the most dramatic prediction of grand unification is
proton decay. This important process which would provide the
window to view physics at truly short distance ($< 10^{-30}$ cm)
and would greatly complement studies of neutrino oscillations in
this regard is yet to be seen. One can, however, argue that the
evidence listed above in favor of supersymmetric grand
unification, based on an effective $G(224)$ or $SO(10)$ symmetry
in 4D, strongly suggests that an upper limit on proton lifetime is
given by 
\[\tau_{\mathrm{proton}} \lsim (\frac{1}{3} -2)\times 10^{34} \mbox{ yrs},\]
with $\bar{\nu} K^{+}$ being the dominant mode, and
quite possibly $\mu^+ K^o$ and $e^+ \pi^o$ being  prominent.
This in turn suggests that an improvement in the current
sensitivity by a factor of five to ten (relative to SuperK) ought
to reveal proton decay. A next-generation megaton-size detector of
the kind being contemplated by the UNO \cite{UNO} and the
Hyperkamiokande \cite{hyperk} proposals would thus be needed to
probe efficiently into the prediction of the supersymmetric
$G(224)/SO(10)$-framework as regards proton decay.

I have discussed in a recent review \cite{JCPICTPtalk} in some
detail the updated results for proton decay in the context of
supersymmetric $SU(5)$, $SO(10)$ and $G(224)$-symmetries by taking
into account (a) the recently improved (and enhanced) matrix
elements as well as short and long-distance renormalization
effects, (b) the dependence of the "standard" d=5 proton-decay
operator on GUT-scale threshold corrections that are restricted by
the requirement of natural coupling unification, and (c) its link
with the masses and the mixings of all fermions including
neutrinos \cite{BPW1}. The latter give rise to a new set of $d=5$
operators, related to the Majorana masses of the RH neutrinos
\cite{BPW2}, which are found to be important. I will present a
summary of the main results in this regard in Sec. 5 and also
comment on the recent works which tend to avoid the standard $d=5$
proton decay operators which generically arise in the context of
supersymmetric grand unification.

In Sec. 2, I discuss the implications of the meeting of the three
gauge couplings in the context of string-unification. In Sec. 3, I
discuss fermion masses and neutrino oscillations within a
predictive framework based on the $G(224)$ or $SO(10)$ symmetry in
4D which lead to predictions of the type shown in
Eq.(\ref{predictions}), and in Sec. 4, I discuss leptogenesis
within the same framework. Results on proton decay which arise
within this framework and within related approaches are summarized
in Sec. 5. Concluding remarks are presented in Sec. 6, where 
the case for building a major underground detector with improved 
sensitivity to detecting proton decay and neutrino oscillations 
is made.
\end{section}

\begin{section}{MSSM Versus String Unifications: 
$G(224)$ Versus $SO(10)$ as Effective Symmetries}
As mentioned in the introduction, the three gauge couplings are found to
meet when they are extrapolated from their values measured at LEP to
higher energies by assuming that the SM is replaced by the minimal
supersymmetric standard model (MSSM) above a threshold of about
$1\ \mathrm{TeV}$ \cite{susyunif}.  The meeting occurs to a very good
approximation, barring a few percent discrepancy which can legitimately
be attributed to GUT-scale threshold corrections.  Their scale of meeting
is given by:
\begin{equation}
M_{X} \approx
2 \times 10^{16}\ \mathrm{GeV}\ \mathrm{(MSSM\ or\ SUSY}\ SU(5)\mathrm{)}
\label{e5}
\end{equation}

This dramatic meeting of the three gauge couplings
provides a strong support for the ideas
of both grand unification and supersymmetry, as being relevant to
physics at short distances $\lsim (10^{16}\ \mathrm{GeV})^{-1}$.

In addition to being needed for achieving coupling unification
there is of course an independent motivation for low-energy
supersymmetry---i.e. for the existence of SUSY partners of the
standard model particles with masses of order 1 TeV.  This is
because it protects the Higgs boson mass from getting large
quantum corrections, which would (otherwise) arise from grand
unification and Planck scale physics.  It thereby provides at
least a technical resolution of the so-called gauge-hierarchy
problem. \emph{In this sense low-energy supersymmetry seems to be
needed for the consistency of the hypothesis of grand
unification.} Supersymmetry is of course also needed for the
consistency of string theory.  It
is fortunate that low-energy supersymmetry can be tested at the
LHC, and possibly at the Tevatron, and the proposed NLC.

The most straightforward interpretation of the observed meeting of the
three gauge couplings and of the scale $M_{X}$, is that a supersymmetric
grand unification symmetry (often called GUT symmetry), like $SU(5)$ or
$SO(10)$, breaks spontaneously at $M_{X}$ into the standard model symmetry
$G(213)$, and that supersymmetry-breaking induces soft masses of order one
TeV.

Even if supersymmetric grand unification may well be a good
effective theory below a certain scale $M \gsim M_X$, it ought to
have its origin within an underlying theory like the string/M
theory.  Such a theory is needed to unify all the forces of nature
including gravity, and to provide a good quantum theory of
gravity.  It is also needed to provide a rationale for the
existence of flavor symmetries (not available within grand
unification), which distinguish between the three families and can
resolve certain naturalness problems including those associated
with inter-family mass hierarchy.  As alluded to in the
introduction, in the context of string or M-theory, an alternative
interpretation of the observed meeting of the gauge couplings is
however possible.  This is because, even if the effective symmetry
in four dimensions emerging from a higher dimensional string
theory is non-simple, like $G(224)$ or even $G(213)$, string
theory can still ensure familiar unification of the gauge
couplings at the string scale. In this case, however, one needs to
account for the small mismatch between the MSSM unification scale
$M_{X}$ (given above), and the string unification scale, given by
\(M_{st}\approx g_{st} \times5.2 \times 10^{17}\ \mathrm{GeV}
\approx 3.6 \times 10^{17}\ \mathrm{GeV}\) (Here we have put
\(\alpha_{st}=\alpha_{\mathrm{GUT}}(\mathrm{MSSM})\approx0.04\))
\cite{Ginspang}. Possible resolutions of this mismatch have been
proposed.  These include:  (i) utilizing the idea of
\emph{string-duality} \cite{WittenDual} which allows a lowering of
$M_{st}$ compared to the value shown above, or alternatively (ii)
the idea of the so-called ``Extended Supersymmetric Standard
Model'' (ESSM) that assumes the existence of two vector-like
families, transforming as \((\mathbf{16}+\overline{\mathbf{16}})\)
of $SO(10)$, with masses of order one TeV \cite{BabuJi}, in
addition to the three chiral families.  The latter leads to a
semi-perturbative unification by raising $\alpha_{\mathrm{GUT}}$
to about 0.25-0.3. Simultaneously, it raises $M_{X}$, in two loop,
to about $(1/2-2)\times10^{17}$ GeV.  (Other mechanisms resolving
the mismatch are reviewed in Ref. \cite{DienesJCP}). In practice,
a combination of the two mechanisms mentioned above may well be
relevant.\footnote{I have in mind the possibility of
string-duality \cite{WittenDual} lowering $M_{st}$ for the case of
semi-perturbative unification in ESSM (for which
$\alpha_{st}\approx 0.25$, and thus, without the use of
string-duality, $M_{st}$ would have been about $10^{18}$ GeV) to a
value of about (1-2)$\times10^{17}$ GeV (say), and
semi-perturbative unification \cite{BabuJi} raising the MSSM value
of $M_{X}$ to about $5\times10^{16}\ \mathrm{GeV} \approx M_{st}$
(1/2 to 1/4) (say).  In this case, an intermediate symmetry like
$G(224)$ emerging at $M_{st}$ would be effective only within the
short gap between $M_{st}$ and $M_{X}$, where it would break into
$G(213)$. Despite this short gap, one would still have the
benefits of $SU(4)$-color that are needed to understand neutrino
masses (see Section 3), and to implement baryogenesis via
leptogenesis.  At the same time, since the gap is so small, the
couplings of $G(224)$, unified at $M_{st}$ would remain
essentially so at $M_{X}$, so as to match with the ``observed''
coupling unification, of the type suggested in Ref.
\cite{BabuJi}.}

While the mismatch can thus quite plausibly be removed for a
non-GUT string-derived symmetry like $G(224)$ or $G(213)$, a GUT
symmetry like $SU(5)$ or $SO(10)$ would have an advantage in this
regard because it would keep the gauge couplings together between
$M_{st}$ and $M_{X}$ (even if $M_{X}\sim M_{st}/20$), and thus not
even encounter the problem of a mismatch between the two scales. A
supersymmetric four dimensional GUT-solution [like $SU(5)$ or
$SO(10)$], however, has a possible disadvantage as well, because it
needs certain color triplets to become superheavy by the so-called
doublet-triplet splitting mechanism in order to avoid the problem of
rapid proton decay.  However, no such mechanism has emerged yet,
in string theory, for the four-dimensional GUT-like solutions
\cite{StringGUT}.\footnote{Some alternative mechanisms for doublet-triplet
splitting, and for suppression of the $d=5$ proton decay operators
have been proposed in the context of higher dimensional theories.
These will be mentioned briefly in Section 5.}

Non-GUT string solutions, based on symmetries like $G(224)$ or $G(2113)$ for
example, have a distinct advantage in this regard, in that the dangerous
color triplets, which would induce rapid proton decay, are often
naturally projected out for such solutions
\cite{stringth, StringG(224),FaraggiHalyo}.  Furthermore, the non-GUT 
solutions
invariably possess new ``flavor'' gauge symmetries, which distinguish
between families and also among members within a family. 
These symmetries are immensely helpful in explaining
qualitatively the observed fermion mass-hierarchy (see e.g.  Ref.
\cite{FaraggiHalyo}) and resolving the so-called naturalness problems of
supersymmetry such as those pertaining to the issues of
squark-degeneracy \cite{FaraggiJCP}, CP violation \cite{BabuJCP} and
quantum gravity-induced rapid proton decay \cite{JCPProton}.

Weighing the advantages and possible disadvantages of both, it
seems hard at present to make a priori a clear choice between a
GUT versus a non-GUT string-solution.  As expressed elsewhere
\cite{JCPRef}, it therefore seems prudent to keep both options
open and pursue their phenomenological consequences.  Given the
advantages of $G(224)$ or $SO(10)$ in understanding the neutrino
masses and implementing leptogenesis (see Sections 3 and 4), I
will thus proceed by assuming that either a suitable four
dimensional $G(224)$-solution with the scale $M_X$ being close to
$M_{st}$ (see footnote 5), or a realistic four-dimensional
$SO(10)$-solution with the desired mechanism for doublet-triplet
splitting, emerges effectively from an underlying string theory,
at the ``conventional'' string-scale \(M_{st} \sim
10^{17}-10^{18}\ \mathrm{GeV}\), and that the $G(224)/SO(10)$
symmetry in turn breaks spontaneously at the conventional
GUT-scale of $M_X\sim 2\times 10^{16}$ GeV (or at $M_X\sim 5\times
10^{16}$ GeV for the case of ESSM, as discussed in footnote 4) to
the standard model symmetry $G(213)$. The extra dimensions of
string/M-theory are assumed to be tiny with sizes $\leq
M_X^{-1}\sim 10^{-30}$ cm, so as not to disturb the successes of
GUT.  In short, I assume that essentially \emph{the conventional
(good old) picture of grand unification, proposed and developed
sometime ago \cite{JCPandAS,GeorgiGlashow,GQW,susyunif}, holds as
a good effective theory above the unification scale $M_X$ and up
to some high scale $M\lsim M_{st}$, with the added presumption
that it may have its origin from the string/M-theory.}\footnote{Alternative
scenarios such as those based on TeV-scale large extra dimensions
\cite{Antoniadis} or string-scale being at a few TeV
\cite{Lykken}, or submillimeter-size even larger extra dimensions
with the fundamental scale of quantum gravity being a few TeV
\cite{submili}, though intriguing, do not seem to provide simple
explanations of these features: (a), (b), and (c). They will be
mentioned briefly in Section 5.2.5.}

 We will see that with the broad assumption mentioned above, an economical
 and predictive framework emerges, which successfully accounts for a host of
 observed phenomena pertaining to the masses and the mixings of all fermions,
including neutrinos, and the baryon asymmetry of the universe.
It also makes some crucial testable predictions for
proton decay.
\end{section}

\begin{section}{Link Between Fermion Masses and Neutrino Oscillations
within a $G(224)/SO(10)$ Framework}

Following Ref.~\cite{BPW1}, I now present a simple and predictive
fermion mass-matrix based on $SO(10)$ or the
$G(224)$-symmetry.\footnote{I will present the Higgs system for
$SO(10)$.  The discussion would remain essentially unaltered if
one uses the corresponding $G(224)$-submultiplets instead.} One
can obtain such a mass mass-matrix for the fermions by utilizing
only the minimal Higgs system that is needed to break the
gauge symmetry $SO(10)$ to \(SU(3)^{c} \times U(1)_{em}\).  It
consists of the set:
\begin{equation}
H_{\mathrm{minimal}} =
\left\{ \mathbf{45_{H}}, \mathbf{16_{H}},
\overline{\mathbf{16}}_{\mathbf{H}}, \mathbf{10_{H}} \right\}
\label{Hmin}
\end{equation}
Of these, the VEV of \(\left\langle \mathbf{45_{H}} \right\rangle
\sim M_{X}\) breaks $SO(10)$ in the B-L direction to \(G(2213) =
SU(2)_{L} \times SU(2)_{R} \times U(1)_{B-L} \times SU(3)^{c}\),
and those of \(\left\langle \mathbf{16_{H}} \right\rangle =
\left\langle \overline{\mathbf{16}}_{\mathbf{H}} \right\rangle\)
along \(\left\langle \tilde{\bar{\nu}}_{RH} \right\rangle\) and
\(\left\langle \tilde{\nu}_{RH} \right\rangle\)\ break $G(2213)$
into the SM symmetry $G(213)$ at the unification-scale $M_{X}$.
Now $G(213)$ breaks at the electroweak scale by the VEV of
\(\left\langle \mathbf{10_{H}} \right\rangle\) to \(SU(3)^{c}
\times U(1)_{em}\).\footnote{Large dimensional tensorial
multiplets of $SO(10)$ like$\mathbf{126_{H}}$,
$\overline{\mathbf{126}}_{H}$, $\mathbf{120_{H}}$, and
$\mathbf{54_{H}}$ are not used for the purpose in part because
they do not seem to arise at least in weakly interacting heterotic
string solutions \cite{DienesMarch}, and in part because they tend
to give too large threshold corrections to $\alpha_{3}(m_{Z})$
(typically exceeding 20\%), which would render observed coupling
unification fortuitous [see e.g. discussions in Appendix D of
Ref.~\cite{BPW1}].}

The question is: can the minimal Higgs system provide a realistic pattern
for fermion masses and mixings?  Now $\mathbf{10_{H}}$ (even several
$\mathbf{10}$'s) 
can not provide certain desirable features ---
i.e. family-antisymmetry and (B-L)-dependence 
in the mass matrices --- which are, however, needed
respectively to suppress $V_{cb}$ while enhancing 
$\theta_{\nu_{\mu}\nu{_{\tau}}}$ on the one hand, and accounting for
features such as \(m_{\mu}^{0} \neq m_{s}^{0}\) on the other hand
(see e.g. 
Ref.~\cite{JCPICTPtalk}, Sec.~5).  
Furthermore, a single $\mathbf{10_{H}}$ cannot generate CKM
mixings.  At the same time, $\mathbf{10_{H}}$ is the only multiplet among
the ones in the minimal Higgs system (Eq. (\ref{Hmin})) which can have cubic
couplings with the matter fermions which are in the $\mathbf{16}$'s.  This
impasse disappears as soon as one allows for not only cubic but also
effective non-renormalizable quartic couplings of the minimal set of
Higgs fields with the fermions.  Such effective couplings can of course
arise quite naturally through exchanges of superheavy states (e.g.
those in the string-tower or those having GUT-scale masses) involving
renormalizable couplings, and/or through quantum gravity.

The $3\times3$ Dirac masses matrices for the four sectors ($u$,
$d$, $l$, $\nu$) proposed in Ref.~\cite{BPW1} are motivated in
part by the group theory of $SO(10)/G(224)$, which severely
restricts the effective cubic and quartic couplings (and thus the
associated mass-patterns), for the minimal Higgs system. They are
also motivated in part by the notion that flavor symmetries
\cite{Flavorsymm} distinguishing between the three families lead
to a hierarchical pattern for the mass matrices (i.e. with the
element ``33'' $\gg$ ``23'' $\gg$ ``22'' $\gg$ ``12'' $\gg$ ``11''
etc.), \emph{so that the lighter family gets its mass primarily
through its mixing with the heavier ones}.  It turns out that the
allowed forms of effective couplings and the corresponding pattern
of mass-matrices, satisfying the constraints of group-theory and
flavor-hierarchy (as above), are rather unique, barring a few
discrete variants.  The mass matrices proposed in Ref.~\cite{BPW1}
are as follows:\footnote{The zeros in ``11'', ``13'', ``31'', and
``22'' elements signify that they are relatively small.  For
instance, the ``22''-elements are set to zero because (restricted
by flavor symmetries, see below), they are meant to be less than
(``23'')(``32'')/``33'' $\sim 10^{-2}$, and thus unimportant for
our purposes \cite{BPW1}.  Likewise, for the other 
``zeros.''} \footnote{A somewhat analogous pattern, also based 
on $SO(10)$, has
been proposed by C. Albright and S. Barr [AB] \cite{AlbrightBarr}. One
major difference between the work of AB and that of BPW
\cite{BPW1} is that the former introduces the so-called
"lop-sided" pattern in which some of the "23" elements are even
greater than the "33" element; in BPW on the otherhand, the
pattern is consistently hierarchical with individual "23" elements
(like $\eta$, $\epsilon$ and $\sigma$) being much smaller in
magnitude than the "33" element of 1.}
\begin{eqnarray}
\label{eq:mat}
\begin{array}{cc}
M_u=\left[
\begin{array}{ccc}
0&\epsilon'&0\\-\epsilon'&0&\sigma+\epsilon\\0&\sigma-\epsilon&1
\end{array}\right]{\cal M}_u^0;&
M_d=\left[
\begin{array}{ccc}
0&\eta'+\epsilon'&0\\ \eta'-\epsilon'&0&\eta+\epsilon\\0&
\eta-\epsilon&1
\end{array}\right]{\cal M}_d^0\\
&\\
M_\nu^D=\left[
\begin{array}{ccc}
0&-3\epsilon'&0\\-3\epsilon'&0&\sigma-3\epsilon\\
0&\sigma+3\epsilon&1\end{array}\right]{\cal M}_u^0;&
M_l=\left[
\begin{array}{ccc}
0&\eta'-3\epsilon'&0\\ \eta'+3\epsilon'&0&\eta-3\epsilon\\0&
\eta+3\epsilon&1
\end{array}\right]{\cal M}_d^0\\
\end{array}
\end{eqnarray}
These matrices are defined in the gauge basis and are multiplied
by $\bar\Psi_L$ on left and $\Psi_R$ on right. For instance, the
row and column indices of $M_u$ are given by $(\bar u_L, \bar c_L,
\bar t_L)$ and $(u_R, c_R, t_R)$ respectively. Note the
group-theoretic up-down and quark-lepton correlations: the same
$\sigma$ occurs in $M_u$ and $M_\nu^D$, and the same $\eta$ occurs
in $M_d$ and $M_l$. It will become clear that the $\epsilon$ and
$\epsilon'$ entries are proportional to B-L and are antisymmetric
in the family space (as shown above). Thus, the same $\epsilon$
and $\epsilon'$ occur in both ($M_u$ and $M_d$) and also in
($M_\nu^D$ and $M_l$), but $\epsilon\rightarrow -3\epsilon$ and
$\epsilon'\rightarrow -3\epsilon'$ as $q\rightarrow l$. Such
correlations result in an enormous reduction of parameters and
thus in increased predictivity. Although the entries $\sigma$,
$\eta$, $\epsilon$, $\eta'$, and $\epsilon'$ will be treated as
parameters, consistent with assignment of flavor-symmetry charges
(see below), we would expect them to be hierarchical with
\((\sigma, \eta, \epsilon) \sim 1/10\) and (\(\eta', \epsilon')\
\sim 10^{-3}-10^{-4}\) (say). Such a hierarchical pattern for the
mass-matrices can be obtained, using a minimal Higgs system ${\bf
45}_H,{\bf 16}_H,{\bf \bar {16}}_H \mbox{ and }{\bf 10}_H $
 and a singlet $S$ of $SO(10)$,
through effective couplings as follows \cite{FN26}:
{\small
\begin{eqnarray}
\lefteqn{\mathcal{L}_{\mathrm{Yuk}} = h_{33}{\bf 16}_3{\bf 16}_3{\bf 10}_H}
\nonumber \\
& & \mbox{} +\left[
h_{23}{\bf 16}_2{\bf 16}_3{\bf 10}_H(S/M)+a_{23}{\bf 16}_2{\bf 16}_3{\bf 10}_H
({\bf 45}_H/M')(S/M)^p+g_{23}{\bf 16}_2{\bf 16}_3{\bf 16}_H^d
({\bf 16}_H/M'')(S/M)^q\right] \nonumber \\
& & + \mbox{}
\left[h_{22}{\bf 16}_2{\bf 16}_2{\bf 10}_H(S/M)^2+g_{22}{\bf 16}_2{\bf 16}_2
{\bf 16}_H^d({\bf 16}_H/M'')(S/M)^{q+1} \right] \nonumber \\
& & \mbox{} +
\left[g_{12}{\bf 16}_1{\bf 16}_2
{\bf 16}_H^d({\bf 16}_H/M'')(S/M)^{q+2}+
a_{12}{\bf 16}_1{\bf 16}_2
{\bf 10}_H({\bf 45}_H/M')(S/M)^{p+2}
\right]
\label{eq:Yuk}
\end{eqnarray}
}
Typically we expect $M'$, $M''$ and $M$ to be of order $M_{\rm string}$
\cite{FN6}. The VEV's of $\langle{\bf 45}_H\rangle$ (along B-L),
$\langle{\bf 16}_H\rangle=\langle{\bf\bar {16}}_H\rangle$ (along standard
model
singlet sneutrino-like component) and of the SO(10)-singlet $\langle S \rangle$
are of the GUT-scale, while those of ${\bf 10}_H$ and of the down type
SU(2)$_L$-doublet component in ${\bf 16}_H$ (denoted by ${\bf 16}_H^d$) are
of the electroweak scale \cite{BPW1,FN7}. Depending upon whether
$M'(M'')\sim M_{\rm GUT}$ or $M_{\rm string}$ (see comment in \cite{FN6}),
the
exponent $p(q)$ is either one or zero \cite{FN8}.

The entries 1 and $\sigma$ arise respectively from $h_{33}$ and
$h_{23}$ couplings, while $\hat\eta\equiv\eta-\sigma$ and $\eta'$
arise respectively from $g_{23}$ and $g_{12}$-couplings. The
(B-L)-dependent antisymmetric entries $\epsilon$ and $\epsilon'$
arise respectively from the $a_{23}$ and $a_{12}$ couplings. This
is because, with $\langle{\bf 45}_H\rangle\propto$ B-L, the
product ${\bf 10}_H\times{\bf 45}_H$ contributes as a {\bf 120},
whose coupling is family-antisymmetric.  Thus, for the minimal
Higgs system (see Eq.~6), (B-L)-dependence can enter only through
family off-diagonal couplings of \(\mathbf{10_{H}} \cdot
\mathbf{45_{H}}\) as in $a_{23}$ and $a_{12}$-terms.  \emph{Thus,
for such a system, the diagonal ``33'' entries are necessarily
(B-L)-independent (as shown in Eq.~(\ref{eq:mat})).  This in turn
makes the relations like \(m_{b}(M_{X}) \approx m_{\tau}\)
(barring corrections of order $\epsilon^{2}$ \cite{BPW1}) robust}.
This feature would, however, be absent if one had used
\(\overline{\mathbf{126}}_{\mathbf{H}}\), whose coupling is
family-symmetric and can give (B-L) dependent contributions to the
``33''-elements.

As alluded to above,
such a hierarchical form of the mass-matrices, with $h_{33}$-term being
dominant, is attributed in part to flavor gauge symmetry(ies) that
distinguishes between the three families \cite{FN9}, and in part to higher
dimensional operators involving for example $\langle{\bf 45}_H\rangle/M'$
or $\langle{\bf 16}_H\rangle/M''$, which are supressed by
$M_{\rm GUT}/M_{\rm string}\sim 1/10$, if $M'$ and/or
$M''\sim M_{\rm string}$.

To discuss the neutrino sector one must specify the Majorana mass-matrix of
the RH neutrinos as well. These arise from the effective couplings of the
form
\cite{FN30}:
\begin{eqnarray}
\label{eq:LMaj}
{\cal L}_{\rm Maj}=f_{ij}{\bf 16}_i{\bf 16}_j{\bf\bar{16}}_H{\bf\bar{16}}_H/M
\end{eqnarray}
where the $f_{ij}$'s include appropriate powers of $\langle S \rangle/M$, in
accord with flavor charge assignments of ${\bf 16}_i$ (see \cite{FN9}). For
the $f_{33}$-term to be leading, we must assign the charge $-a$ to
${\bf\bar{16}}_H$. This leads to a hierarchical form for the Majorana
mass-matrix
\cite{BPW1}:
\begin{eqnarray}
\label{eq:MajMM}
M_R^\nu=\left[
\begin{array}{ccc}
x&0&z\\0&0&y\\z&y&1
\end{array}
\right]M_R
\end{eqnarray}
Following the flavor-charge assignments given in \cite{FN9}, we
expect \(|y|\sim \langle S/M\rangle\sim 1/10\),
\(|z|\sim (\langle S/M\rangle)^2\sim 10^{-2}(1 \mbox{ to }
1/2)\),  \(|x|\sim (\langle S/M\rangle)^4\sim (10^{-4}$-$10^{-5})\) (say).
The ``22'' element (not shown) is $\sim (\langle S/M\rangle)^2$ and its
magnitude is
taken to be $< |y^2/3|$, while the ``12'' element (not shown) is
$\sim (\langle S/M\rangle)^3$. We expect
\begin{equation}
M_{R} =
f_{33} \langle \overline{\mathbf{16}}_{\mathbf{H}}
\rangle^{2}/M_{\mathrm{string}} \approx 10^{15}\ \mathrm{GeV} (1/2 - 2) \label{mag22}
\end{equation}
where we have put
\( \langle \overline{\mathbf{16}}_{\mathbf{H}} \rangle \approx M_{X}
\approx 2 \times 10^{16}\ \mathrm{GeV}\),
\(M_{\mathrm{string}} \approx 4 \times 10^{17}\ \mathrm{GeV}\)
\cite{Ginspang}, and \(f_{33} \approx 1\), and
have allowed for an uncertainty by a factor of 2 in the estimate around a centrally expected value of about $10^{15} \mathrm{GeV}$. Allowing for 2-3 family-mixing in the Dirac and the Majorana sectors as in
Eqs. 7 and 10, the seesaw mechanism leads to \cite{BPW1}:
\begin{equation}
m(\nu_{3}) \approx B \frac{m(\nu_{\mathrm{Dirac}}^{\tau})^{2}}{M_{R}} \label{seesaw}
\end{equation}
The quantity $B$ represents the effect of 2-3 mixing and is given
by \(B = (\sigma + 3\epsilon)(\sigma + 3\epsilon - 2y)/y^{2}\)
(see Eq.~(24) of Ref.~\cite{BPW1}).  Thus $B$ is fully calculable
within the model once the parameters $\sigma$, $\eta$, $\epsilon$,
and $y$ are determined in terms of inputs involving some quark and
lepton masses (as noted below).  In this way, one obtains $B
\approx (2.9 \pm 0.5)$.  The Dirac mass of the tau-neutrino is
obtained by using the $SU(4)$-color relation (see Eq. (2)):
\(m(\nu_{\mathrm{Dirac}}^{\tau}) \approx m_{\mathrm{top}}(M_{X})
\approx 120\ \mathrm{GeV}\). One thus obtains from Eq. (12):
\begin{equation}
m(\nu_{3}) \approx \frac{(2.9)(120\ \mathrm{GeV})^{2}}{10^{15}\ \mathrm{GeV}}
(1/2 - 2) \approx (1/24\ \mathrm{eV})(1/2 - 2) \label{seesaw2}
\end{equation}
Noting that for hierarchical entries --- i.e. for ($\sigma$, $\epsilon$,
and $y$) \(\sim 1/10\) --- one naturally obtains a hierarchical spectrum
of neutrino-masses:
\(m(\nu_{1}) \lsim m(\nu_{2}) \sim (1/10)m(\nu_{3})\),
we thus get:
\begin{equation}
\left[ \sqrt{\Delta m_{23}^{2}} \right]_{\mathrm{Theory}}
\approx m(\nu_{3}) \approx (1/24\ \mathrm{eV})(1/2 - 2) \label{delta}
\end{equation}
This agrees remarkably well with the SuperK value of
\((\sqrt{\Delta m_{23}^{2}})_{\mathrm{SK}} (\approx 1/20 \mbox{
eV})\), which lies in the range of nearly (1/15 to 1/30) eV.  As
mentioned in the introduction, the success of this prediction
provides clear support for (i) the existence of $\nu_{R}$, (ii)
the notion of $SU(4)$-color symmetry that gives
$m(\nu_{\mathrm{Dirac}}^{\tau})$, (iii) the SUSY unification-scale
that gives $M_{R}$, and (iv) the seesaw mechanism.

We note that alternative symmetries such as $SU(5)$ would have no
compelling reason to introduce the $\nu_{R}$'s.  Even if one did
introduce $\nu_{R}^{i}$ by hand, there would be no symmetry to relate the
Dirac mass of $\nu_{\tau}$ to the top quark mass.  Thus
$m(\nu_{\mathrm{Dirac}}^{\tau})$ would be an arbitrary parameter in $SU(5)$.
Furthermore, without B-L as a local symmetry, the Majorana masses of the
RH neutrinos, which are singlets of $SU(5)$, can naturally be of order
string scale \(\sim 4 \times 10^{17}\ \mathrm{GeV}\) (say).  That would,
however, give too small a mass for $m(\nu_{3})$ ($ < 10^{-4}\ \mathrm{eV}$)
compared to the SuperK value.

Other effective symmetries such as $[SU(3)]^{3}]$ \cite{su33} 
and \(SU(2)_{L} \times SU(2)_{R}\times U(1)_{B-L}\times SU(3)^{C}\) \cite{2213}
would give $\nu_{R}$ and B-L as a local symmetry, but not the desired
$SU(4)$-color mass-relations: \(m(\nu_{\mathrm{Dirac}}^{\tau})
\approx m_{t}(M_{X})\) and \(m_{b}(M_{X}) \approx m_{\tau}\). Flip
\(SU(5) \times U(1)\) \cite{flip} on the other hand would yield the desired
features for the neutrino-system, but not the $b$-$\tau$ mass
relation. Thus, combined with the observed $b/\tau$ mass-ratio,
the SuperK data on atmospheric neutrino oscillation seems to
clearly select out the effective symmetry in 4D being either
$G(224)$ or $SO(10)$, as opposed to the other alternatives
mentioned above. \emph{It is in this sense that the neutrinos, 
by virtue of their tiny masses, provide crucial information
on the unification-scale as
well as on the nature of the unification-symmetry in 4D, as
alluded to in the introduction}.

Ignoring possible phases in the parameters and thus the source of CP violation
for a
moment, as was done in Ref. \cite{BPW1}, the parameters $(\sigma,\eta,
\epsilon, \epsilon',\eta', {\cal M}_u^0, {\cal M}_D^0,\mbox{ and } y)$ can be
determined by using, 
for example, $m_t^{\rm phys}=174$ GeV, $m_c(m_c)=1.37$ GeV,
$m_S(1\mbox{ GeV})=110$-116 MeV, $m_u(1\mbox{ GeV})=6$ MeV, the observed
masses of $e$, $\mu$, and $\tau$ and
$m(\nu_2)/m(\nu_3)\approx 1/(7\pm 1)$ (as suggested by a combination
of atmospheric \cite{sk}
and solar neutrino data \cite{sno}, the latter corresponding to
the LMA MSW solution, see
below) as inputs. One is thus led, {\it for this CP conserving case}, to the
 following fit for the parameters, and the
associated predictions \cite{BPW1}. [In this fit, we leave the small quantities
$x$ and $z$ in $M_R^\nu$ undetermined and proceed by assuming that they have the
magnitudes suggested by flavor symmetries
(i.e., $x\sim (10^{-4}$-$10^{-5})$ and $z\sim 10^{-2}$(1 to 1/2)
(see remarks below Eq. (\ref{eq:MajMM}))]:
\begin{eqnarray}
\label{eq:fit}
\begin{array}{c}
\sigma\approx 0.110, \quad \eta\approx 0.151, \quad \epsilon\approx -0.095,
 \quad |\eta'|\approx 4.4 \times 10^{-3},\\
\begin{array}{cc}
\epsilon'\approx 2\times 10^{-4},& {\cal M}_u^0\approx
m_t(M_X)\approx 120 \mbox{ GeV}, \\{\cal M}^0_D\approx
m_b(M_X)\approx 1.5 \mbox{ GeV}, & y\approx -(1/17).
\end{array}
\end{array}
\end{eqnarray}
These in turn lead to the following predictions for the quarks and
light neutrinos \cite{BPW1}:\footnote{These predictions are based on the fact
that the pattern given in Eq.~7 leads to
\(m_{b}(M_{X}) \approx m_{\tau}(1-8\epsilon^{2})\).
They also reflect the recent trend in the atmospheric and solar
neutrino data which suggests
\(m(\nu_{2})/m(\nu_{3}) \approx 1/7\).}
\begin{eqnarray}
\label{eq:pred}
\begin{array}{l}
m_b(m_b)\approx(4.7\mbox{-}4.9)\mbox{ GeV},\\
m(\nu_{3}) \approx (1/24\ \mathrm{eV})(1/2 - 2),\\
V_{cb}\approx\left|\sqrt{\frac{m_s}{m_b}}\left|\frac{\eta+\epsilon}
{\eta-\epsilon}\right|^{1/2}- \sqrt{\frac{m_c}{m_t}}\left|\frac{\sigma
+\epsilon}{\sigma-\epsilon}\right|^{1/2}\right|\approx 0.044,\\
\left\{ \begin{array}{l}
\theta^{\rm osc}_{\nu_\mu\nu_\tau}\approx\left|\sqrt{\frac{m_\mu}{m_\tau}}
\left|\frac{\eta-3\epsilon}{\eta+3\epsilon}\right|^{1/2}+
\sqrt{\frac{m_{\nu_2}}{m_{\nu_3}}}\right|\approx |0.437+(0.378\pm 0.03)|,\\
\mbox{Thus, } \sin^2 2\theta^{\rm osc}_{\nu_\mu\nu_\tau}\approx
0.99,\ \  (\mbox{for } m(\nu_{2})/m(\nu_{3}) \approx 1/7),\\
\end{array}\right.\\
V_{us}\approx \left|\sqrt{\frac{m_d}{m_s}}-\sqrt{\frac{m_u}{m_c}}\right|
\approx 0.20,\\
\left|\frac{V_{ub}}{V_{cb}} \right|\approx \sqrt{\frac{m_u}{m_c}}\approx
0.07,\\
m_d(\mbox{1 GeV})\approx \mbox{8 MeV},\\
\theta^{\rm osc}_{\nu_e\nu_\mu}\approx 0.06  \mbox{ (ignoring
non-seesaw contributions; see, however, remarks below)}
\end{array}
\end{eqnarray}

The Majorana masses of the RH neutrinos ($N_i$) are given
by:\footnote{The range in $M_{3}$ and $M_{2}$ is constrained by the
values of $m(\nu_{3})$ and $m(\nu_{2})$ suggested by the atmospheric and
solar neutrino data.}
\begin{eqnarray}
\label{eq:MajM}
M_{3}& \approx & M_R\approx 10^{15}\mbox{ GeV (1/2-1)},\nonumber\\
M_{2}& \approx & |y^2|M_{3}\approx \mbox{(2.5$\times 10^{12}$ GeV)(1/2-1)},\\
M_{1}& \approx & |x-z^2|M_{3}\sim (1/2\mbox{-}2)10^{-5}M_{3}\sim
\mbox{$10^{10}$ GeV(1/4-2)}.\nonumber
\end{eqnarray}

Note that we necessarily have a hierarchical spectrum for the
light as well as the heavy neutrinos (see discussions below on
$m_{\nu_1}$). Leaving out the $\nu_e$-$\nu_\mu$ oscillation angle
for a moment, it seems remarkable that the first seven predictions
in Eq. (\ref{eq:pred}) agree with observations, to within 10\%.
Particularly intriguing is the (B-L)-dependent
{\it group-theoretic correlation}
between the contribution from the first term in $V_{cb}$ and that
in $\theta^{\rm osc}_{\nu_\mu\nu_\tau}$, which explains
simultaneously why one is small ($V_{cb}$) and the other is large
($\theta^{\rm osc}_{\nu_\mu\nu_\tau}$) \cite{newFN36}. That in
turn provides some degree of confidence in the gross structure of
the mass-matrices.

As regards $\nu_e$-$\nu_\mu$ and $\nu_e$-$\nu_\tau$ oscillations,
the standard seesaw mechanism would typically lead to rather small
angles as in Eq. (\ref{eq:pred}), within the framework presented
above \cite{BPW1}. It has, however, been noted recently
\cite{JCPICTPtalk} that small intrinsic (non-seesaw) masses $\sim
10^{-3}$ eV of the LH neutrinos can arise quite plausibly through
higher dimensional operators of the form \cite{FN32}:
\(W_{12}\supset \kappa_{12}{\bf 16}_1{\bf 16}_2{\bf 16}_H{\bf
16}_H{\bf 10}_H {\bf 10}_H/M_{\rm eff}^3\), without involving the
standard seesaw mechanism \cite{seesaw}. One can verify that such
a term would lead to an intrinsic Majorana mixing mass term of the
form \(m_{12}^0\nu_L^e\nu_L^\mu\), with a strength given by
\(m_{12}^0\approx \kappa_{12}\langle{\bf 16}_H \rangle^2(175\mbox{
GeV})^2/ M_{\rm eff}^3\sim (1.5\mbox{-}6)\times 10^{-3}\) eV, for
\(\langle{\bf 16}_H \rangle\approx (1\mbox{-}2)M_{\rm GUT}\) and
$\kappa_{12}\sim 1$, if \(M_{\rm eff}\sim M_{\rm GUT}\approx
2\times 10^{16}\) GeV \cite{FN33}. Such an intrinsic Majorana 
$\nu_e\nu_\mu$ mixing mass $\sim $ few$\times 10^{-3}$ eV, though
small compared to $m(\nu_3)$, is still much larger than what one
would generically get for the corresponding term from the standard
seesaw mechanism [as in Ref. \cite{BPW1}]. Now, the diagonal
($\nu_L^\mu\nu_L^\mu$) mass-term, arising from the standard seesaw
mechanism is expected to be \(\sim (3-8) \times 10^{-3}\) eV for
a natural value of 
$|y|\approx 1/20$-1/15, say \cite{BPW1}. Thus, taking the net
values of \(m_{22}^{0} \approx 7 \times 10^{-3}\) eV, \(m_{12}^0
\approx 3 \times 10^{-3}\) eV as above and \(m_{11}^0 \ll
10^{-3}\) eV, which are all plausible, we obtain  \(m_{\nu_{2}}
\approx 7 \times 10^{-3}\) eV, \(m_{\nu_{1}} \sim \mbox{(1 to
few)} \times 10^{-3}\) eV, so that \(\Delta m^2_{12}\approx 5
\times 10^{-5}\ \mathrm{eV}^{2}\) and \(\sin^{2}
2\theta_{12}^{\mathrm{osc}} \approx 0.6-0.7\). These go well with
the LMA MSW solution of the solar neutrino problem.

In summary, \emph{the intrinsic non-seesaw contribution} to the
Majorana masses of the LH neutrinos (neglected in making the
predictions of Eq. (\ref{eq:pred})) can plausibly have the right
magnitude for $\nu_e$-$\nu_\mu$ mixing so as to lead to the LMA
solution within the $G(224)/SO(10)$-framework, without upsetting
the successes of the first seven predictions in Eq.
(\ref{eq:pred}). [In contrast to the near maximality of the
$\nu_\mu$-$\nu_\tau$ oscillation angle, however, which emerges as
a compelling prediction of the framework \cite{BPW1}, the LMA
solution, as obtained above, should, be regarded only as a
consistent possibility, rather than as a compelling prediction,
within this framework.]

It is worth noting at this point that in a theory leading to Majorana 
masses of the LH neutrinos as above, \emph{one would of course expect the 
neutrinoless double beta decay process (like $n+n\rightarrow ppe^- 
e^-$), satisfying $|\Delta L|=2$ and $|\Delta B|=0$, to occur at 
some level.} Search for this process is most important because it 
directly tests a fundamental conservation law and can shed light 
on the Majorana nature of the neutrinos, as well as on certain CP
violating phases in the neutrino-system (assuming that the process is
dominated by neutrino-exchange). The crucial parameter which 
controls the strength of this process is given by $m_{ee} \equiv 
|\sum_i m_{\nu_i} U_{e i}^2|$. With a non-seesaw contribution leading 
to \(m_{\nu_1}\sim\mbox{ few }\times 10^{-3}\mbox{ eV}\), 
\(m_{\nu_2}\approx\mbox 7\times 10^{-3}\mbox{ eV}\), 
\(\sin^2 2\theta_{12} \approx 0.6-0.7\), and an expected value for 
\(\sin \theta_{13} \sim m^{0}_{13} / m^{0}_{33} \sim (1-5)\times 10^{-3} 
\mbox{ eV }/(5\times 10^{-2}\mbox{ eV }) \sim (0.02-0.1)\), one would 
expect \(m_{ee}\approx (1-5)\times 10^{-3}\mbox{ eV}\). Such a strength, 
though compatible with current limits \cite{Vogel}, 
would be accessible if the current 
sensitivity is improved by about a factor of 50--100. Improving the 
sensitivity to this level would certainly be most desirable. 

I would now like to turn to a discussion of leptogenesis within the
$G(224)/SO(10)$-framework for fermion masses and mixings presented
above.
Before discussing leptogenesis, we need to discuss, however, the origin of CP
violation within the same framework. The discussion
so far
has ignored, for the sake of simplicity, possible CP violating phases in the
parameters ($\sigma$, $\eta$, $\epsilon$, $\eta'$, $\epsilon'$,
$\zeta_{22}^{u,d}$, $y$, $z$, and $x$) of the Dirac and Majorana mass matrices
[Eqs. (\ref{eq:mat}), and (\ref{eq:MajMM})]. In general, however, these
parameters can and generically will have phases \cite{FN34}. Some combinations
of these phases enter into the CKM matrix and define the Wolfenstein parameters
$\rho_W$ and $\eta_W$ \cite{Wolfenstein}, which in turn induce CP violation by
utilizing the standard model interactions. As observed in
Ref.~\cite{BabuJCPCascais},
an additional and potentially important source of CP
and flavor violations (as in $K^0\leftrightarrow\bar K^0$,
$B_{d,s}\leftrightarrow\bar B_{d,s}$, $b\rightarrow s\bar s s$, etc.
transitions)
arise in the model through supersymmetry \cite{FN36}, involving squark and
gluino loops (box and penguin), simply because of the embedding of MSSM within
a string-unified $G(224)$ or $SO(10)$-theory near the GUT-scale, and the
assumption that primordial SUSY-breaking occurs near the string scale
($M_{\rm string}>M_{\rm GUT}$) \cite{FN37}.
 It is shown in \cite{BabuJCPCascais}
that complexification of the parameters ($\sigma$, $\eta$, $\epsilon$,
$\eta'$, $\epsilon'$, etc.), through introduction of phases $\sim 1/20$-1/2
(say) in them, can still preserve the successes of the predictions as regards
fermion masses and neutrino oscillations shown in Eq. (\ref{eq:pred}), as
long as one maintains nearly the magnitudes of the real parts of the
parameters and especially their relative signs as obtained in Ref.~\cite{BPW1}
and shown in Eq. (\ref{eq:fit}) \cite{FN38}. Such a picture is also in accord
with the observed features of CP and flavor violations in $\epsilon_K$,
$\Delta m_{Bd}$, and asymmetry parameter in
 $B_d\rightarrow J/\Psi+K_s$, while predicting observable new effects in
 processes such as $B_s\rightarrow \bar B_s$ and $B_d\rightarrow \Phi+K_s$
 \cite{BabuJCPCascais}.

We therefore proceed to discuss leptogenesis concretely within the
framework presented above by adopting the Dirac and Majorana
fermion mass matrices as shown in Eqs. (\ref{eq:mat}) and
(\ref{eq:MajMM}) and assuming that the parameters appearing in
these matrices can have natural phases $\sim 1/20$-1/2 (say) with
either sign up to addition of $\pm \pi$, {\it while their real
parts have the relative signs and nearly the magnitudes given in
Eq. (\ref{eq:pred}).}
\end{section}

\begin{section}{Leptogenesis}
Finally, the observed matter-antimatter asymmetry of the universe
provides an additional important clue to physics at truly short
distances. This issue has taken a new turn since the discovery of
the non-perturbative electroweak sphaleron effects \cite{KRS},
which violate B+L but conserve B-L. These remain in thermal
equilibrium in the temperature range of 200 GeV to about $10^{12}$
GeV. As a result, they efficiently erase any pre-existing
baryon/lepton asymmetry that satisfies $\Delta (\mathrm{B+L}) \neq
0$, but $\Delta (\mathrm{B-L}) = 0$. This is one reason why
standard GUT-baryogenesis satisfying $\Delta (\mathrm{B-L}) = 0$
(as in minimal SU(5)) becomes irrelevant to the observed baryon
asymmetry of the universe\footnote{Standard GUT-baryogenesis
involving decays of X and Y gauge bosons (with $M_X \sim 10^{16}$
GeV) and/or of superheavy Higgs bosons is hard to realize anyway
within a plausible inflationary scenario satisfying the
gravitino-constraint [see e.g. E. W. Kolb and M. S. Turner, "The
Early Universe", Addison-Wesley, 1990].}. On the other hand,
purely electroweak baryogenesis based on the sphaleron effects -
although a priori an interesting possibility - appears to be
excluded for the case of the standard model without supersymmetry,
and highly constrained as regards the available parameter space
for the case of the supersymmetric standard model, owing to LEP
lower limit on Higgs mass $\geq 114$ GeV. As a
result, in the presence of electroweak
 sphalerons, baryogenesis via leptogenesis \cite{Vanagidalepto} appears to be
an attractive and promising mechanism to generate the observed baryon
asymmetry of the universe.

To discuss leptogenesis concretely within the G(224)/SO(10) -
framework presented above, I follow the discussion of
Ref.\cite{JCPlepto} and first consider the case of thermal
leptogenesis. In the context of an inflationary scenario \cite{d},
with a plausible reheat temperature $T_{RH}\sim (1 \mbox{ to
few})\times 10^{9}$ GeV (say), one can avoid the well known
gravitino problem if $m_{3/2}\sim (1\mbox{ to }2)$ TeV
\cite{gravitino1} and yet produce the lightest heavy neutrino
$N_1$ efficiently from the thermal bath if $M_{1}\sim\mbox{(3 to
5)} \times 10^{9}$ GeV (say), in accord with Eq. (\ref{eq:MajM})
[$N_2$ and $N_3$ are of course too heavy to be produced at $T\sim
T_{RH}$]. Given lepton number (and B-L) violation occurring
through the Majorana mass of $N_1$, and C and CP violating phases
in the Dirac and/or Majorana fermion mass-matrices as mentioned
above, the out-of-equilibrium decays of $N_1$ (produced from the
thermal bath) into $l+H$ and $\bar l+\bar H$ and into the
corresponding SUSY modes $\tilde l+\tilde H$ and $\bar {\tilde
l}+\bar {\tilde H}$ would produce a B-L violating lepton
asymmetry; so also would the decays of $\tilde N_1$ and
$\bar{\tilde{N_1}}$. Part of this asymmetry would of course be
washed out due to inverse decays and lepton number violating
2$\leftrightarrow$2-scatterings. I will assume this commonly
adopted mechanism for the so-called thermal leptogenesis (At the
end, I will consider an interesting alternative that
would involve non-thermal leptogenesis). This mechanism has been
extended to incorporate supersymmetry by several authors (see
e.g., \cite{Campbell,Covi,Plumacher}). The net lepton asymmetry of
the universe [$Y_L\equiv(n_L-n_{\bar L})/s$] arising from decays
of $N_1$ into
 $l+H$ and $\bar l+\bar H$ as well as into the corresponding SUSY modes
($\tilde l+\tilde H$ and $\bar {\tilde l}+\bar {\tilde H}$) and likewise from
$(\tilde N_1, \bar{\tilde{N_1}})$-decays \cite{Campbell,Covi,Plumacher} is
given by:
\begin{eqnarray}
\label{eq:YL} Y_L=\kappa\epsilon_1\left(\frac{n_{N_1}+n_{\tilde
N_1}+ n_{\bar{\tilde{N}}_1}}{s}\right)\approx \kappa\epsilon_1/g^*
\end{eqnarray}
where $\epsilon_1$ is the lepton-asymmetry produced per $N_1$ (or
$(\tilde N_1+\bar{\tilde{N_1}})$-pair) decay (see below), and
$\kappa$ is the efficiency or damping factor that represents the
washout effects mentioned above (thus $\kappa$ incorporates the
extent of departure from thermal equilibrium in $N_1$-decays; such
a departure is needed to realize lepton asymmetry).\footnote{The efficiency 
factor mentioned above, is often expressed in terms of the
parameter $K\equiv [\Gamma(N_1)/2H]_{T=M_1}$ \cite{d}. Assuming initial
thermal abundance for $N_1$, $\kappa$ is normalized so that it is 1 if
$N_1$'s decay fully out of equilibrium corresponding to $K\ll 1$
(in practise, this actually requires $K<0.1$).} The parameter
$g^*\approx 228$ is the number of light degrees of freedom in
MSSM.

The lepton asymmetry $Y_L$ is converted to baryon asymmetry, by the sphaleron
effects, which is given by:
\begin{eqnarray}
\label{eq:YB}
Y_B=\frac{n_B-n_{\bar B}}{s}=C\,Y_L,
\end{eqnarray}
where, for MSSM, $C \approx -1/3$.
Taking into account the interference between the tree and loop-diagrams for
the decays of $N_1\rightarrow lH$ and $\bar l\bar H$ (and likewise for
$N_1\rightarrow \tilde l\tilde H$ and $\bar{\tilde l}\bar{\tilde H}$ modes
and also for $\tilde N_1$ and $\bar{\tilde{N_1}}$-decays), the CP violating
lepton asymmetry parameter in each of the four channels (see e.g.,
\cite{Covi} and \cite{Plumacher}) is given by
\begin{eqnarray}
\label{eq:epsilon1}
\epsilon_1=\frac{1}{8\pi v^2(M_D^\dagger M_D)_{11}}\sum_{j=2,3}
{\rm Im} \left[(M_D^\dagger M_D)_{j1} \right]^2 f(M_j^2/M_1^2)
\end{eqnarray}
where $M_D$ is the Dirac neutrino mass matrix evaluated in a basis
in which the Majorana mass matrix of the RH neutrinos $M_R^\nu$
[see Eq. (\ref{eq:MajMM})] is diagonal, $v=(\mbox{174 GeV})
\sin\beta$ and the function $f\approx -3(M_1/M_j)$, for the case
of SUSY with $M_j\gg M_1$.

Including inverse decays as
well as $\Delta L\neq 0$-scatterings in the Boltzmann equations, a recent
analysis \cite{Bari} shows that in the relevant parameter-range of interest
to us (see below), the efficiency factor (for the SUSY case) is given by
\cite{newFN50}:
\begin{eqnarray}
\label{eq:d}
\kappa\approx (0.7\times 10^{-4})({\rm eV}/\tilde{m_1})
\end{eqnarray}
where $\tilde{m_1}$ is an effective mass parameter (related to $K$
\cite{newFN51}), and is given by \cite{m_tilde}:
\begin{eqnarray}
\label{eq:k}
\tilde{m_1}\equiv (m^\dagger_Dm_D)_{11}/M_1.
\end{eqnarray}
Eq. (\ref{eq:d}) should hold to better than 20\%
(say), when $\tilde m_1\gg 5\times 10^{-4}$ eV \cite{Bari}
(This applies well to our case, see below).

I proceed to make
numerical estimates of the lepton-asymmetry
by taking the magnitudes and the relative signs of the real parts of the
parameters ($\sigma$, $\eta$, $\epsilon$, $\eta'$, $\epsilon'$, and $y$)
approximately the same as in Eq.~(\ref{eq:fit}), but allowing in general
for natural phases in them (as in \cite{BabuJCPCascais}).

In the evaluation of the lepton asymmetry, I allow for small ``31''
and "13" entries in $M_\nu^D$, denoted by $\zeta_{31}$ and
$\zeta_{13}$ respectively, which are not exhibited in Eq.
(\ref{eq:mat}). Following assignment of flavor-charges \cite{FN9},
these are expected to be of order $(1/200)(1 \mbox{
}\mathrm{to}\mbox{ } 1/2)$ (say). As such, they have no noticeable
effects on fermion masses and mixings discussed above, but they
can be relevant to lepton asymmetry.

Using the values of the parameters ($\sigma$, $\epsilon$,
$\epsilon'$, $y$, and $z$) determined from our
consideration of fermion masses (see Eq.~(\ref{eq:fit})) and the
expected magnitudes of $\zeta_{31}$ and $z$, one obtains the following
estimates (see Ref. \cite{JCPlepto} for details):
\begin{eqnarray}
\frac{\left( M_D^\dagger M_D\right)_{11}}{\left({\cal M}_u^0\right)^2}
& = &
|3\epsilon'-z(\sigma-3\epsilon)|^2+\left|\zeta_{31}-z\right|^2 \nonumber \\
& \approx & 2.5\times 10^{-5}(1/4\mbox{ to }1/6) \label{eq:terms11} \\
\epsilon_{1} & \approx & \frac{1}{8\pi}\left(\frac{{\cal
M}_u^0}{v}\right)^2
|(\sigma+3\epsilon)-y|^2\sin\left(2\phi_{21}\right)
\left(\frac{-3M_1}{M_2}\right)\approx -2.0\times
10^{-6} \sin\left(2\phi_{21}\right) \nonumber\\
\label{eq:epsilon2}
\end{eqnarray}
where,
\(\phi_{21} = \mathrm{arg}[(\zeta_{31}-z)(\sigma^{*}+3\epsilon^{*}-y^{*})]+
(\phi_1-\phi_2)\).
Here $(\phi_1 -\phi_2)$ is a phase angle that arises from
diagonalization of the Majorana mass matrix $M_R^\nu$ (see \cite{JCPlepto}).
\emph{The effective phase $\phi_{21}$ thus depends upon phases in both 
the Dirac and the Majorana mass matrices.}
In writing Eq. (\ref{eq:epsilon2}),
we have put \((\mathcal{M}_u^{0}/v)^2 \approx 1/2\),
\(|\sigma+3\epsilon-y| \approx 0.13\)
(see Eq. (\ref{eq:fit}) and Ref. \cite{FN38}), and
for concreteness (for the present case of thermal leptogenesis)
$M_1\approx 4\times 10^9$ GeV and $M_2\approx 2\times 10^{12}
\mbox{ GeV}$ [see Eq.~(\ref{eq:MajM})]. Since $|\zeta_{31}|$ and $|z|$ are
each expected to be of $(1/200) (1 to 1/2)$ (say) by flavor symmetry, we have
allowed for a possible mild cancellation between their contributions to
$|\zeta_{31} -z|$ by putting $|\zeta_{31} -z|\approx (1/200) (1/2 -1/5)$.
The parameter
$\tilde{m_1}$, given by Eq.~(\ref{eq:k}), turns out to be (approximately)
proportional to
$|\zeta_{31}-z|^2$ [see Eq.~(\ref{eq:terms11})].
It is given by:
\begin{eqnarray}
\label{eq:k2}
\tilde{m_1}\approx |\zeta_{31}-z|^2 ({\cal M}_u^0)^2/M_1\approx
(1.9\times 10^{-2}\mbox{ eV})(\mbox{1 to 1/6})
\bigg(\frac{4\times10^9 \mathrm{GeV}}{M_1}\bigg)
\end{eqnarray}
where, as before, we have put 
$|\zeta_{31}-z|\approx (1/200)(1/2\mbox{ to }1/5)$. The
corresponding efficiency factor $\kappa$ [given by Eq.
(\ref{eq:d})], lepton and baryon-asymmetries $Y_L$ and $Y_B$
[given by Eqs. (\ref{eq:YL}) and (\ref{eq:YB})] and the
requirement on the phase-parameter $\phi_{21}$ are listed in
Table~\ref{tab:phase}.
\begin{table}
\begin{center}
\begin{tabular}{|c|c|c|c|c|}
\hline
&\multicolumn{3}{c|}{$|\zeta_{31}-z|$}\\
\hhline{|~|-|-|-|}
&(1/200)(1/3)&(1/200)(1/4)&(1/200)(1/5)\\ \hline\hline
$\tilde{m}_{1}$(eV) & $0.83\times 10^{-2}$ & $0.47\times 10^{-2}$ &
$0.30\times 10^{-2}$ \\ \hline
$\kappa$ & 1/73 & 1/39&1/24 \\ \hline
$Y_L/\sin(2\phi_{21})$ & $-11.8\times 10^{-11}$ & $-22.4\times
10^{-11}$ & $-36\times 10^{-11}$ \\ \hline
$Y_B/\sin(2\phi_{21})$ & $4\times 10^{-11}$ & $7.5\times 10^{-11}$ & $12\times
10^{-11}$ \\ \hline
$\phi_{21}$ & $\sim\pi/4$ & $\sim\pi/12-\pi/4$ & $\sim\pi/18-\pi/4$ \\ \hline
\end{tabular}
\end{center}
\caption{Baryon Asymmetry for the Case of Thermal Leptogenesis}
\label{tab:phase}
\end{table}

The constraint on $\phi_{21}$ is obtained from considerations of
Big-Bang nucleosynthesis, which requires \(3.7 \times 10^{-11}
\lesssim (Y_B)_{BBN} \lesssim 9 \times 10^{-11}\) \cite{BBN};
this is consistent with the CMB data \cite{CMB}, which
suggests somewhat higher values of \((Y_{B})_{CMB} \approx
(7-10)\times 10^{-11}\) (say). We see that the first case
\(|\zeta_{31}-z| \approx 1/200(1/3)\) leads to a baryon asymmetry
$Y_B$ that is on the low side of the BBN-data, even for a maximal
\(\sin(2\phi_{21})\approx 1\). The other cases with
$|\zeta_{31}-z|\approx (1/200)(1/4\mbox{ to }1/5)$, which
 are of course
perfectly plausible, lead to the desired magnitude of the baryon asymmetry
for natural values of the phase parameter
\(\phi_{21}\sim (\pi/18\mbox{ to }\pi/4)\). We see that, for the thermal case,
the CMB data would suggest somewhat smaller values of
\(|\zeta_{31} -z| \sim 10^{-3}\).
This constraint would be eliminated for the case of non-thermal leptogenesis.

We next consider briefly the scenario of non-thermal leptogenesis
\cite{JeannerotNonThermal,OtherNonThermal}. In this case the
inflaton is assumed to decay, following the inflationary epoch, 
directly into a pair of heavy RH neutrinos (or sneutrinos).
These in turn decay into $l+H$ and $\bar{l} +\bar{H}$ as well as
into the corresponding SUSY modes, and thereby produce lepton asymmetry, during
the process of reheating. It turns out that this scenario
goes well with the fermion mass-pattern of Sec. 2 [in particular see Eq.
\eqref{eq:MajM}] and the observed baryon asymmetry, provided
$2M_2>m_{\rm infl}>2M_1$, so that the inflaton
decays into $2N_1$ rather than into $2N_2$ (contrast this from the case
proposed in Ref.~\cite{JeannerotNonThermal}). In this case, the reheating
temperature ($T_{\rm RH}$) is found to be much less than
$M_1\sim 10^{10}$ GeV (see below); thereby
(a) the gravitino constraint is satisfied quite easily, even for a rather
low gravitino-mass $\sim 200$ GeV (unlike in the thermal case); at the same
time (b) while $N_1$'s are produced non-thermally (and copiously) through
inflaton decay, they remain out of equilibrium and the wash out process
involving inverse decays and $\Delta L\neq 0$-scatterings are ineffective,
so that the efficiency factor $\kappa$ is 1.

To see how the non-thermal case can arise naturally, we recall
that the VEV's of the Higgs fields $\Phi=(1,2,4)_H$ and $\bar
\Phi=(1,2,\bar 4)_H$ have been utilized to (i) break SU(2)$_R$ and
B-L so that G(224) breaks to the SM symmetry~\cite{JCPandAS}, and
simultaneously (ii) to give Majorana masses to the RH neutrinos
via the coupling in Eq.~\eqref{eq:LMaj} (see e.g., Ref.~\cite{BPW1}; 
for SO(10), $\bar \Phi$ and $\Phi$ would be in ${\bf
16}_H$ and $\bar{\bf 16}_H$ respectively). It is attractive to
assume that the same $\Phi$ and $\bar\Phi$ (in fact their
$\tilde{\nu_{\rm RH}}$ and $\bar{\tilde{\nu}}_{\rm
RH}$-components), which acquire GUT-scale VEV's, also drive
inflation \cite{JeannerotNonThermal}. In this case the inflaton
would naturally couple to a pair of RH neutrinos by the coupling
of Eq.~\eqref{eq:LMaj}. To implement hybrid inflation in this
context, let us assume following Ref.~\cite{JeannerotNonThermal},
an effective superpotential \(W_{\rm eff}^{\rm infl}=\lambda
S(\bar\Phi\Phi-M^2)+\mbox{(non-ren. terms)}\), where $S$ is a
singlet field \cite{M2Vevs}. It has been shown in 
Ref.~\cite{JeannerotNonThermal} that in this case a flat potential with
a radiatively generated slope can arise so as to implement
inflation, with $G(224)$ broken during the inflationary epoch to
the SM symmetry. The inflaton is made of two complex scalar fields
(i.e., \(\theta=(\delta\tilde\nu_H^C+\delta\tilde{\bar \nu}_H^C)/
\sqrt{2}\) that represents the fluctuations of the Higgs fields
around the SUSY minimum, and the singlet S). Each of these have a
mass $m_{\rm infl}=\sqrt{2}\lambda M$, where \(M=\langle\mbox{(1,
2, 4)}_H\rangle\approx 2\times 10^{16}\) GeV and a width
\(\Gamma_{\rm
infl}=\Gamma(\theta\rightarrow\Psi_{\nu_H}\Psi_{\nu_H}) =
\Gamma(S\rightarrow\tilde\nu_H\tilde\nu_H) \approx
[1/(8\pi)](M_1/M)^2{m}_{\rm infl}\) so that
\begin{equation}
T_{\rm RH}\approx (1/7)(\Gamma_{\rm infl}M_{\rm Pl})^{1/2} \approx (1/7)(M_1/M)
[m_{\rm infl}M_{\rm Pl}/(8\pi)]^{1/2}
\end{equation}
For concreteness, take \cite{NonThermalRelaxedGravitino}
\(M_2\approx 2\times 10^{12}\) GeV, \(M_1\approx 2\times 10^{10}\) GeV (1 to 2)
[in accord with Eq. \eqref{eq:MajM}], and $\lambda\approx 10^{-4}$, so
that ${m}_{\rm infl}\approx 3\times 10^{12}$ GeV.
We then get:
\(T_{\rm RH}\approx (1.7\times 10^8\mbox{ GeV})\mbox{(1 to 2)}\),
and thus (see e.g., Sec. 8 of
Ref. \cite{d}):
\begin{eqnarray}\label{YB2}
(Y_B)_{Non-Thermal}&\approx& -(Y_L/3) \nonumber\\
&\approx& (-1/3) [(n_{N_1}+n_{\tilde N_1}+n_{\bar{\tilde{N}}_1})/s]\epsilon_1
\nonumber\\
&\approx& (-1/3)[(3/2)(T_{\rm RH}/m_{\rm infl})\epsilon_1] \nonumber\\
&\approx& (30\times 10^{-11})(\sin 2\phi_{21})\mbox{(1 to 2)}^2
\end{eqnarray}
Here we have used Eq. \eqref{eq:epsilon2} for $\epsilon_1$ with
appropriate $(M_1/M_2)$, as above. Setting $M_1\approx 2\times
10^{10}$ for concreteness, we see that $Y_B$ obtained above agrees
with the (nearly central) observed value of \(\langle
Y_B\rangle^{\rm central}_{\rm BBN(CMB)} \approx (6(8.6)) \times
10^{-11}\), again for a natural value of the phase parameter
$\phi_{21}\approx \pi/30 (\pi/20)$. As mentioned above, one
possible advantage of the non-thermal over the thermal case is
that the gravitino-constraint can be met rather easily, in the
case of the former (because $T_{RH}$ is rather low
$\sim 10^{8}$ GeV), whereas for the thermal case there is a
significant constraint on the lowering of the $T_{RH}$ (so as to
satisfy the gravitino-constraint) vis a vis a raising of $M_1 \sim
T_{RH}$ so as to have sufficient baryon asymmetry (note that
$\epsilon_1 \propto M_1$, see Eq.~(\ref{eq:epsilon2})).
Furthermore, for the non-thermal case, the dependence of $Y_B$ on
the parameter $|\zeta_{31} -z|^2$ (which arises through $\kappa$
and $\tilde{m}_1$ in the thermal case, see Eqs. (\ref{eq:d}),
(\ref{eq:k}), and (\ref{eq:terms11})) is largely eliminated. Thus
the expected magnitude of $Y_{B}$ (Eq.~(\ref{YB2}) holds without
a significant constraint on $|\zeta_{31} - z|$ (in contrast to the
thermal case).

To conclude this part, I have considered two alternative scenarios
(thermal as well as non-thermal) for inflation and leptogenesis.
\emph{We see that the $G(224)/SO(10)$ framework provides a simple and
unified description of not only fermion masses and neutrino
oscillations (consistent with maximal atmospheric and large solar
neutrino oscillation angles) but also of baryogenesis via
leptogenesis, for the thermal as well as non-thermal cases, in
accord with the gravitino constraint}. Each of the features
--- (a) the existence of the right-handed neutrinos, (b) B-L local
symmetry, (c) quark-lepton unification through SU(4)-color, (d)
the seesaw mechanism, and (e) the magnitude of the supersymmetric
unification-scale
--- plays a crucial role in
realizing this unified and successful description. These features
in turn point to the relevance of either the G(224) or the SO(10)
symmetry being effective between the string and the GUT scales, in
four dimensions.  While the observed magnitude of the baryon
asymmetry seems to emerge naturally from within the framework,
understanding its observed sign (and thus the relevant CP
violating phases) remains a challenging task.
\end{section}



\begin{section}{Proton Decay: The Hallmark of Grand Unification}
\begin{subsection}{Preliminaries}
Turning to proton decay, I present now the reason why the unification
framework based on SUSY $SO(10)$ or SUSY $G(224)$, together with the
understanding of fermion masses and mixings discussed above, strongly
suggest that proton decay should be imminent.

In supersymmetric unified theories there are in general three
distinct mechanisms for proton decay - two realized sometime ago
and one rather recently. Briefly, they are:
\begin{enumerate}
\item The familiar $d=6$ operators mediated by X and Y gauge
bosons of $SU(5)$ or $SO(10)$ (Fig.~1).  These lead to
$e^{+}\pi^{0}$ as the dominant mode. \item The ``standard'' $d=5$
operators \cite{Sakai} (Fig.~2) which arise through exchanges of
color triplet Higgsinos which are in the \(\mathbf{5}_{\mathrm{H}}
+\bar{\mathbf{5}}_{\mathrm{H}}\) of $SU(5)$ or $10_{\mathrm{H}}$
of $SO(10)$.  In the presence of these operators, one crucially
needs, for consistency with the empirical lower limit on proton
lifetime, a suitable doublet-triplet splitting mechanism which
assigns GUT-scale masses to the color triplets in the
$\mathbf{10}_{\mathrm{H}}$ of $SO(10)$ while keeping the
electroweak doublets light (see e.g.\ Ref.~\cite{BPW1} for
discussion of such mechanisms and relevant references). Following
the constraints of Bose symmetry, color antisymmetry and
hierarchical Yukawa couplings, these standard $d=5$ operators lead
to dominant $\bar{\nu}K^{+}$ and comparable $\bar{\nu}\pi^{+}$
modes, but in all cases to highly suppressed $e^{+}\pi^{0}$,
$e^{+}K^{0}$ and even $\mu^{+}K^{0}$ modes. 
\item The ``new'' $d=5$ operators \cite{BPW2} which arise (see Fig.~3) through
exchanges of color triplet Higgsinos in the Higgs multiplets like
\(\mathbf{16}_{\rm H} + \overline{\mathbf{16}}_{\mathrm{H}}\) of
$SO(10)$, which are used to give superheavy Majorana masses to the
RH neutrinos. These operators
generically arise through the joint effects of (a) the couplings
as in Eq.~(\ref{eq:LMaj}) which assign superheavy Majorana masses
to the RH neutrinos, and (b) the couplings of the form \(g_{ij}
\mathbf{16}_{i} \mathbf{16}_{j} \mathbf{16}_{\mathrm{H}}
\mathbf{16}_{\mathrm{H}}/M\) as in Eq.~(\ref{eq:Yuk}) which are
needed to generate CKM mixings (see Sec.~3).  Thus these new $d=5$
operators are directly linked not only to the masses and mixings
of quarks and leptons, but also to the Majorana masses of the RH
neutrinos.
\end{enumerate}

The contributions of these three operators to proton decay has been
considered in detail in Ref.~\cite{JCPICTPtalk}, which provides an
update in this regard of the results of Ref.~\cite{BPW1}.  Here,
I will highlight only the main ingredients that enter into the
calculations of proton decay rate, based on the three contributions
listed above, and then present a summary of the main results.  The
reader is referred to these two references for a more detailed
presentation and explanations.

Relative to other analyses, the study of proton decay carried
out in Refs.~\cite{BPW1} and \cite{JCPICTPtalk} have the following
distinctive features:

\textbf{(i) Link with Fermion Masses:} It systematically takes into
account the link that exists between the $d=5$ proton decay
operators and the masses and mixings of all fermions including
neutrinos, within a realistic $G(224)/SO(10)$-framework,
as discussed in Sec.~3.

\textbf{(ii) Inclusion of the standard and the new $d=5$
operators:}  In particular, it includes contributions from both
the standard and the new $d=5$ operators (Fig.~3), related to the
Majorana masses of the RH neutrinos.  These latter, invariably
omitted in the literature, are found to be generally as important
as the standard ones.

\textbf{(iii) Restricting GUT-scale Threshold Corrections:} The
study restricts GUT-scale threshold corrections to
$\alpha_{3}(m_{Z})$ so as to be in accord with the demand of
``natural'' coupling unification.  This restriction is especially
important for SUSY $SO(10)$, for which, following the mechanism of
doublet-triplet splitting (see Appendix of Ref.~\cite{BPW1}), the
standard $d=5$ operators become inversely proportional to an
effective mass-scale given by \(M_{\mathrm{eff}} \equiv
\left[\lambda \langle \mathbf{45}_{\mathrm{H}} \rangle
\right]^{2}/ M_{10'} \sim M_{X}^{2}/M_{10'}\), rather than to the
physical masses of the color-triplets in the
$\mathbf{10}_{\mathrm{H}}$ of $SO(10)$.  Here $M_{10'}$ represents
the mass of $\mathbf{10}'_{\mathrm{H}}$, that enters into the D-T
splitting mechanism through an effective coupling \(\lambda
\mathbf{10}_{\mathrm{H}} \mathbf{45}_{\mathrm{H}}
\mathbf{10}'_{\mathrm{H}}\) in the superpotential.  Now, $M_{10'}$
can be naturally suppressed compared to $M_{X}$ owing to flavor
symmetries, and thus $M_\mathrm{eff}$ can be correspondingly
larger than $M_{X}$ by even two to three orders of
magnitude.\footnote{It should be noted that $M_{\mathrm{eff}}$
does not represent the physical masses of the color-triplets or of
the other particles in the theory.  It is simply a parameter of
order $M_{X}^{2}/M_{10'}$ that is relevant to proton decay.  Thus
values of $M_{\mathrm{eff}}$, close to or even exceeding the
Planck scale, does not in any way imply large corrections from
quantum gravity.} \footnote{Accompanying the suppression due to
$M_{\mathrm{eff}}$, it turns out that the standard $d=5$ operators
for $SO(10)$ possess an intrinsic enhancement as well, compared to
those for $SU(5)$, primarily due to correlations between the
Yukawa couplings in the up and down sectors in $SO(10)$.  The
standard $d=5$ amplitude for proton decay in $SO(10)$ is thus
based on these two opposing effects --- suppression through
$M_{\mathrm{eff}}$ and enhancement through the Yukawa couplings
\cite{BPW1}.}

Although $M_{\mathrm{eff}}$ can far exceed $M_X$, it still gets
bounded from above by demanding that coupling unification, as
observed,\footnote{For instance, in the absence of GUT-scale
threshold corrections, the MSSM value of $\alpha_3(m_Z)_{MSSM}$,
assuming coupling unification, is given by
$\alpha_3(m_Z)_{MSSM}^\circ=0.125\pm 0.13$ \cite{susyunif}, which
is about 5-8\% higher than the observed value:
$\alpha_3(m_Z)_{MSSM}^\circ=0.118 \pm 0.003$. We demand that this
discrepancy should be accounted for accurately by a net
\emph{negative} contribution from D-T splitting and from ``other''
GUT-scale threshold corrections, without involving large
cancellations. That in fact does happen for the minimal Higgs
system $(\mathbf{45}, \mathbf{16}, \overline{\mathbf{16}})$ (see
Ref.~\cite{BPW1}).} should emerge as a natural prediction of the
theory as opposed to being fortuitous. That in turn requires that
there be no large (unpredicted) cancellation between GUT-scale
threshold corrections to the gauge couplings that arise from
splittings within different multiplets as well as from Planck
scale physics.  Following this point of view, we have argued (see
Ref.~\cite{BPW1}) that the net ``other'' threshold corrections to
$\alpha_3(m_Z)$ arising from the Higgs and the gauge multiplets
should be negative, but conservatively and quite plausibly no more
than about 10\%, at the electroweak scale. Such a requirement is
in fact found to be well satisfied not only in magnitude but also
in sign by the minimal Higgs system consisting of
($\mathbf{45}_{\mathrm{H}}$, $\mathbf{16}_{\mathrm{H}}$
$\overline{\mathbf{16}}_{\mathrm{H}}$, and
$\mathbf{10}_{\mathrm{H}}$) \cite{BPW1}. This in turn restricts
how big can be the threshold corrections to $\alpha_3(m_Z)$ that
arise from (D-T) splitting (which is positive).  Since the latter
turns out to be proportional to $\ln(M_{\rm
eff}\cos\gamma/M_{X})$, we thus obtain an upper limit on $M_{\rm
eff}\cos\gamma$, where $\cos\gamma\approx(\tan\beta)/(m_t/m_b)$.
An upper limit on $M_{\rm eff}\cos\gamma$ thus provides an upper
limit on $M_{\rm eff}$ which is inversely proportional to
$\tan{\beta}$. In this way, our demand of natural coupling
unification, together with the simplest model of D-T splitting,
introduces an upper limit on $M_{\mathrm{eff}}$ given by
\(M_{\mathrm{eff}} \leq 2.7 \times 10^{18} \mbox{ GeV} (3/\tan
\beta)\) for the case of MSSM embedded in $SO(10)$.  This in turn
introduces an implicit dependence on $\tan\beta$ into the lower
limit of the SO(10)-amplitude---i.e. $\widehat{A}(SO(10))\propto
1/M_{\rm eff} \geq$ [(a quantity) $\propto \tan\beta$].  These
considerations are reflected in the results given below. [More
details can be found in Ref.\cite{BPW1} and \cite{JCPICTPtalk}].

\textbf{(iv) Allowing for the ESSM Extension of MSSM:} The case of
the extended supersymmetric standard model (ESSM), briefly alluded
to in Sec.~2, is an interesting variant of MSSM, which can be
especially relevant to a host of observable phenomena, including
(a) proton decay, (b) possible departure of muon ($g-2$) from the SM prediction
\cite{BabuJCPg-2}, and (c) a lowering of the LEP neutrino-counting
from the SM value of 3 \cite{BabuJCPNuTeV}.  Briefly
speaking, ESSM introduces an extra pair of vectorlike families
transforming as \(\mathbf{16} + \overline{\mathbf{16}}\) of
$SO(10)$, having masses of order 1~TeV
\cite{BabuJi,BabuPatiStrem}. Adding such complete
$SO(10)$-multiplets would of course preserve coupling unification.
From the point of view of adding extra families, ESSM seems to be
the minimal and also the maximal extension of the MSSM, that is
allowed in that it is compatible with 
(a) precision electroweak tests, as well as (b) a
semi-perturbative as opposed to non-perturbative gauge coupling
unification \cite{BabuJi,KoldaRussell}.\footnote{For instance,
addition of \emph{two} pairs of vector-like families at the
TeV-scale, to the three chiral families, would cause gauge
couplings to become non-perturbative below the unification scale.}
\emph{The existence of two extra vector-like families of quarks
and leptons can of course be tested at the LHC.}

Theoretical motivations for the case of ESSM arise on several
grounds: (a) it provides a better chance for stabilizing the
dilaton by having a semi-perturbative value for
\(\alpha_{\mathrm{unif}} \approx 0.35-0.3\) \cite{BabuJi}, in
contrast to a very weak value of 0.04 for MSSM; (b) owing to
increased two-loop effects \cite{BabuJi,KoldaRussell}, it raises
the unification scale $M_{X}$ to \((1/2 - 2) \times 10^{17}\) GeV
and thereby considerably reduces the problem of a mismatch
\cite{DienesJCP} between the MSSM and the string unification
scales (see Section~2); (c) it lowers the GUT-prediction for
$\alpha_{3}(m_{Z})$ to (0.112--0.118) (in absence of
unification-scale threshold corrections), which is in better
agreement with the data than the corresponding value of
(0.125--0.113) for MSSM; and (d) it provides a simple reason for
inter-family mass-hierarchy \cite{BabuJi,BabuPatiStrem}. In this
sense, ESSM, though less economical than MSSM, offers some
distinct advantages.

In the present context, because of raising of $M_{X}$ and
lowering of $\alpha_{3}(m_{Z})$, ESSM naturally
enhances the GUT-prediction for proton lifetime, in full accord
with the data \cite{SKlimit}.  Specifically, for ESSM, one
obtains: \(M_{\mathrm{eff}} \leq (6 \times 10^{18}\mbox{
GeV})(30/\tan \beta)\) \cite{BPW1,JCPICTPtalk}.

As a result, in contrast to MSSM, ESSM can allow for larger values of
$\tan \beta$ (like 10), or lighter squark masses ($\sim 1\mbox{ TeV}$)
without needing large threshold corrections, and simultaneously without
conflicting with the limit on proton lifetime (see below).
\end{subsection}

\begin{subsection}{Proton Decay Rate}
Some of the original references on contributions of the standard
$d=5$ operators to proton decay may be found in
\cite{DimopRabyWilczek, Ellis, NathChemArno,Hisano, BabuBarr,
BPW1, JCPICTPtalk, LucasRaby, Murayama}. I now specify some of the
parameters involving the matrix element, renormalization effects
and the spectrum of the SUSY partners of the SM particles that are
relevant to calculations of proton decay rate.

The hadronic matrix element is defined by
\(\beta_{H}u_{L}(\vec{k}) \equiv \epsilon_{\alpha \beta \gamma}
\langle 0 \left| (d_{L}^{\alpha} u_{L}^{\beta})
u_{L}^{\gamma} \right| p,\vec{k} \rangle \). A recent improved
lattice calculation yields \(\beta_H \approx 0.014 \mbox{
GeV}^{3}\) \cite{Aoki} (whose systematic errors that may arise
from scaling violations and quenching are hard to estimate).  We
will take as a conservative, but plausible, range for $\beta_{H}$
to be given by \((0.014 \mbox{ GeV}^{3})(1/2-2)\). $A_{S}$ denotes
the short distance renormalization effect for the $d=5$ operator
which arises owing to extrapolation between GUT and SUSY-breaking
scales \cite{Ellis, Hisano, Turznyski}. The average value of
\(A_{S} = 0.67\), given in Ref.~\cite{Hisano} for \(m_{t} = 100
\mbox{ GeV}\), has been used in most early estimates.  For \(m_{t}
= 175 \mbox{ GeV}\), a recent estimate yields: \(A_{S} \approx
0.93 \mbox{ to } 1.2\) \cite{Turznyski}. Conservatively, I would
use \(A_{S} = 0.93\); this would enhance the rate by a factor of
two compared with previous estimates. $A_{L}$ denotes the
long-distance renormalization effect of the $d=6$ operator due to
QCD interaction that arises due to extrapolation between the SUSY
breaking scale and 1 GeV \cite{Ellis}. Using the two-loop
expression for $A_{L}$ \cite{Arafune}, together with the two-loop
value of $\alpha_{3}$, Babu and I find: \(A_{L} \approx 0.32\).
This by itself would also increase the rate by a factor of
\((0.32/0.22)^{2} \approx 2\), compared to the previous estimates
\cite{Ellis, NathChemArno,Hisano, BabuBarr, BPW1}.  Including the
enhancements in both $A_{S}$ and $A_{L}$, we thus see that the net
increase in the proton decay rate solely due to new evaluation of
renormalization effects is nearly a factor of four, compared to
the previous estimates (including that in Ref.~\cite{BPW1}).

In Ref.~\cite{BPW1}, guided by the demand of naturalness (i.e. the
absence of excessive fine tuning), in obtaining the Higgs boson
mass, squark masses were assumed to lie in the range of \(1 \mbox{
TeV}(1/\sqrt{2} - \sqrt{2})\), so that \(m_{\bar{q}} \lsim 1.4
\mbox{ TeV}\).  Work, based on the notion of focus point
supersymmetry however suggests that squarks may be quite a bit
heavier without conflicting with the demands of naturalness
\cite{Feng+}.  In the interest of obtaining a conservative upper
limit on proton lifetime, we will therefore allow squark masses to
be as heavy as about \(2.4 \mbox{ TeV}\).

Allowing for plausible and rather generous uncertainties in the matrix
element and the spectrum I take:
\begin{eqnarray}
\beta_{H} & = & (0.014 \mbox{ GeV}^{3})(1/2 - 2), \nonumber \\
m_{\bar{q}} \approx m_{\bar{l}} & \approx & 1.2 \mbox{ TeV}(1/2 - 2),
\nonumber \\
m_{\bar{w}}/m_{\bar{q}} & = & 1/6(1/2 - 2), \nonumber \\
M_{H_c} (\mbox{minimal SU(5)}) &\leq& 10^{16}\mbox{ GeV},
\nonumber
\\
A_{L} & \approx & 0.32, \nonumber \\
A_{S} & \approx &
\begin{array}{lcr}0.93 & \mbox{and} & \tan \beta \geq 3\end{array}.
\label{new28}
\end{eqnarray}

For evaluation of the strengths of the $d=6$ operators, generated
by exchanges of $X$ and $Y$ gauge bosons, for the cases of SUSY
$SO(10)$ or $SU(5)$ with MSSM spectrum, we take:\footnote{For the
central value of $\alpha_{H}$, I take the value quoted in
Ref.~\cite{Aoki} and allow for an uncertainty by a factor of two
either way around this central value.}
\begin{displaymath}
\left.
\begin{array}{ccc}
M_{X} \approx  M_{Y} & \approx & 10^{16}\mbox{ GeV}(1 \pm 25\%) \\
\alpha_{i}(GUT) & \approx & 0.04
\end{array}
\right\}  (\mbox{for MSSM})
\end{displaymath}
\begin{equation}
\alpha_{H} =  0.015\mbox{ GeV}^{3} (1/2 - 2)
\label{new29}
\end{equation}

Before presenting the theoretical predictions, I note the
following experimental results on inverse proton decay rates
provided by the SuperK studies \cite{SKLimitepi, SKlimit}:
\begin{eqnarray}
\Gamma^{-1}(p \rightarrow e^{+}\pi^{0})_{\mathrm{expt}} &
\gsim & 6 \times 10^{33} \mbox{ yrs} \nonumber \\
\left[ \sum_{l} \Gamma(p \rightarrow \bar{\nu}_{l}K^{+})
\right]_{\mathrm{expt}}^{-1} & \gsim & 1.9 \times 10^{33} \mbox{ yrs}
\label{new30}
\end{eqnarray}

Before the theoretical predictions for proton decay can be given,
a few comments are in order.

\begin{enumerate} \item I present the
results separately for the standard $d=5$ and the new $d=5$
operators, allowing for both the MSSM and the ESSM alternatives.
(The contributions of the new $d=5$ operators are in fact the
same for these two alternatives.)  Although the proton decay
amplitude receives contributions from both the standard and the
new operators, in practice, the standard d=5 operators dominate
over the new ones for the case of MSSM in the parameter-range of
interest that corresponds to predicted proton lifetimes in the
upper end, while the new operators dominate over the standard ones
for the case of ESSM, in the same range.  (This may be inferred
from the results listed below.) Thus, in practice, it suffices to
consider the contributions of the standard and the new operators
separately. \item In evaluating the contributions of the new $d=5$
operators to proton decay, allowance is made for the fact that for
the $f_{ij}$ couplings (see Eq.~(9)), there are two possible
$SO(10)$-contractions (leading to a $\mathbf{45}$ or a
$\mathbf{1}$) of the pair \(\mathbf{16}_{i}
\overline{\mathbf{16}}_{H}\), both of which contribute to the
Majorana masses of the $\nu_{R}$s, but only the contraction via
the $\mathbf{45}$ contributes to proton decay. In the presence of
non-perturbative quantum gravity one would in general expect both
contractions to be present having comparable strengths.  For
example, the couplings of the $\mathbf{45}$s lying in the
string-tower or possibly below the string scale, and likewise of
the singlets, to the \(\mathbf{16}_{i}
\overline{\mathbf{16}}_{H}\) pair would respectively generate the
two contractions.  Allowing for a difference between the relevant
projection factors for $\nu_{R}$-masses versus proton decay
operator, we set \((f_{ij})_{p} \equiv (f_{ij})_{\nu}K\), where
$(f_{ij})_{\nu}$ defined in Sec.~3 directly yields
$\nu_{R}$-masses and $K$ is a relative factor of order
unity.\footnote{Thus the new set of proton decay operators become
proportional to \((f_{ij})_{\nu} g_{kl}K \langle
\overline{\mathbf{16}}_{H} \rangle \langle \mathbf{16}_{H} \rangle
/ (M^{2})(M_{16})\) where \(M \approx M_{\mathrm{st}}\) and
\(M_{16}(\sim M_{\mathrm{GUT}})\) is the mass of the
$\mathbf{16}_{H}$ (see Ref.~\cite{JCPICTPtalk} for a discussion
limiting the strength of this operator).}  As a plausible range,
we take $K \approx 1/5 - 2$ (say), where $K=1/5$ seems to be a
conservative value on the low side that would correspond to proton
lifetimes near the upper end.
\end{enumerate}

The theoretical predictions for proton decay for the case of the 
minimal supersymmetric SU(5) model, and the supersymmetric SO(10) 
and G(224)-models developed in Secs. 3 and 4, are summarized below. 
They are based on (a) the
items (i)--(iv) listed above, (b) the two comments mentioned above, and
(c) the values of the relevant parameters listed in
Eqns.~(\ref{new28}) and (\ref{new29}).\footnote{In obtaining the rate 
for the $e^+\pi^o$-mode
induced by the d=6 operator, we have used the net renormalization factor 
$A_R \approx2.5$ representing long and short-distance effects\cite{renorm6}
and the chiral lagrangian parameters -- D and F as in Ref.
\cite{chirallag}}

\begin{center}
\textbf{A Summary of Results on Proton Decay and Discussions}
\end{center}

{\footnotesize
\begin{eqnarray}
\left. \frac{\mbox{Min. SUSY $SU(5)$}}{\mbox{MSSM (std. $d=5$)}} \right\}
\begin{array}{l}\Gamma^{-1}(p \rightarrow \bar{\nu}K^{+}) \end{array} & \leq &
\begin{array}{lr}
\begin{array}{l} 1.2 \times 10^{31} \mbox{ yrs} \end{array} &
\left( \begin{array}{c} \mbox{Excluded by} \\ \mbox{SuperK} \end{array} \right)
\end{array} \label{new31}\\ \nonumber\\
\left. \frac{\mbox{SUSY $SO(10)$}}{\mbox{MSSM (std. $d=5$)}} \right\}
\begin{array}{l} \Gamma^{-1}(p \rightarrow \bar{\nu}K^{+}) \end{array} & \leq &
\begin{array}{lr}
\begin{array}{l} 1 \times 10^{33} \mbox{ yrs} \end{array} &
\left( \begin{array}{c} \mbox{Tightly constrained} \\
\mbox{by SuperK} \end{array} \right)
\end{array} \label{new32} \\ \nonumber\\
\left. \frac{\mbox{SUSY $SO(10)$}}{\mbox{ESSM (std. $d=5$)}} \right\}
\begin{array}{l}
\Gamma^{-1}(p \rightarrow \bar{\nu}K^{+})_{\mbox{Med.}} \\
\Gamma^{-1}(p \rightarrow \bar{\nu} K^{+})
\end{array}
&
\begin{array}{c} \approx \\ \lsim \end{array}
&
\begin{array}{lr}
\begin{array}{l}
(\mbox{1--10}) \times 10^{33} \mbox{ yrs} \\ 10^{35} \mbox{ yrs}
\end{array}
&
\left( \begin{array}{c} \mbox{Fully SuperK} \\ \mbox{Compatible}
\end{array} \right)
\end{array} \label{new33} \\ \nonumber\\
\left. \frac{\mbox{SUSY $G(224)/SO(10)$}}{\mbox{MSSM or ESSM (new $d=5$)}} \right\}
\begin{array}{l}
\Gamma^{-1}(p \rightarrow \bar{\nu}K^{+}) \\
B(p \rightarrow \mu^{+} K^{0})
\end{array}
&
\begin{array}{c} \lsim \\ \approx \end{array}
&
\begin{array}{lr}
\begin{array}{l}
2 \times 10^{34} \mbox{ yrs} \\ (1-50)\%
\end{array}
&
\left( \begin{array}{c} \mbox{Fully Compatible} \\ \mbox{with SuperK}
\end{array} \right)
\end{array} \label{new34} \\ \nonumber\\
\left. \frac{\mbox{SUSY $SU(5)$ or $SO(10)$}}{\mbox{MSSM ($d=6$)}} \right\}
\begin{array}{l} \Gamma^{-1}(p \rightarrow e^{+} \pi^{0}) \end{array} & \approx
& \begin{array}{lr} \begin{array}{l} 10^{34.9 \pm 1} \mbox{ yrs}
\end{array} & \left( \begin{array}{c} \mbox{Fully Compatible} \\
\mbox{with SuperK}
\end{array} \right)
\end{array} \label{new35}
\end{eqnarray}}

It should be stressed that the upper limits on proton lifetimes
given above are quite conservative in that they are obtained
(especially for the top two cases) by stretching the uncertainties
in the matrix element and the SUSY spectra as given in
Eq.~(\ref{new28}) to their extremes so as to prolong proton
lifetimes.  In reality, the lifetimes should be shorter than the
upper limits quoted above.  With this in mind, the following
comments are in order:
\begin{enumerate}
\item By comparing the upper limit given in Eq.~(\ref{new31}) with
the experimental lower limit (Eq.~(\ref{new30})), we see that the
\emph{minimal} SUSY $SU(5)$ with the conventional MSSM spectrum is
clearly excluded by a large margin by proton decay searches.  This
is in full agreement with the conclusion reached by other authors
(see e.g. Ref.~\cite{Murayama})\footnote{See, however, Refs~\cite{BajcPerezSenja} and \cite{EmmanuelWies}, where attempts are made to save minimal SUSY SU(5) by a set of scenarios, which seems (to me) contrived. These include a judicious choice of sfermion mixings, higher dimensional operators and squarks of first two families having masses of order 10 TeV.}.  We have of course noted in
Sec.~3 that SUSY $SU(5)$ does not go well with neutrino
oscillations observed at SuperK. \item By comparing
Eq.~(\ref{new32}) with the empirical lower limit
(Eq.~(\ref{new30})), we see that the case of MSSM embedded in
$SO(10)$ is already tightly constrained to the point of being
disfavored by the limit on proton lifetime.  The constraint is of
course augmented by our requirement of \emph{natural coupling
unification}, which prohibits accidental large cancellation
between different threshold corrections.\footnote{For instance,
had we allowed the ``other'' GUT-scale threshold corrections (in
our case, those arising from $\mathbf{45}_{H}$, $\mathbf{16}_{H}$,
$\overline{\mathbf{16}}_{H}$ and the gauge multiplet, see
Refs.~\cite{BPW1, JCPICTPtalk}) to $\alpha_{3}(m_{Z})$ to be negative in sign and
large as about 15\% (rather than 10\%), as some authors do
\cite{LucasRaby}, the upper limit on proton lifetime would have been higher
by about a factor of 5, compared to Eq.~(\ref{new32}).} On the
positive side, improvement in the current limit by even a factor
of 2--3 (say) ought to reveal proton decay.  Otherwise the case of
MSSM embedded in $SO(10)$ would be clearly excluded. \item In
contrast to the case of MSSM, that of ESSM embedded in $SO(10)$
(see Eq.~(\ref{new33})) is fully compatible with the SuperK limit.
In this case, \(\Gamma_{\mathrm{Med}}^{-1}(p \rightarrow
\bar{\nu}K^{+}) \approx 10^{33} - 10^{34} \mbox{ yrs}\), given in
Eq.~(\ref{new33}), corresponds to the parameters involving the
SUSY spectrum and the matrix element $\beta_{H}$ being in the
\emph{median range}, close to their central values (see
Eq.~(\ref{new28})). In short, confining to the standard operators
only, if ESSM represents low energy physics and if $\tan \beta$ is
rather small (3 to 5 say), \emph{we do not have to stretch the
uncertainties in the SUSY spectrum and the matrix elements to
their extreme values in order to understand why proton decay has
not been seen as yet, and still can be optimistic that it ought to
be discovered in the near future with a lifetime $\lsim 10^{34}$
yrs.}\footnote{The results on proton lifetimes for a wide
variation of the parameters for the case of MSSM and ESSM embedded
in $SO(10)$ are listed in Tables 1 and 2 of
Ref.~\cite{JCPICTPtalk}.} \item We see from Eq.~(\ref{new34}) that
the contribution of the new operators related to the Majorana
masses of the RH neutrinos (Fig.~3) (which is the same for MSSM
and ESSM and is independent of $\tan \beta$) is fully compatible
with the SuperK limit.  These operators can quite naturally lead
to proton lifetimes in the range of $10^{33}-10^{34}$ yrs with an
upper limit of about \(2 \times 10^{34}\) yrs.

It should be remarked that if in the unlikely event all the
parameters ($\beta_{H}$, $m_{\tilde{W}}/m_{\tilde{q}}$, and
$m_{\tilde{q}}$, etc.) happen to be closer to their extreme values
(see Eq.~(\ref{new28})) so as to extend proton lifetime, the
standard $d=5$ operators for the case of ESSM embedded in $SO(10)$
would lead to lifetimes as long as about $10^{35}$ years (see
Eq.~(\ref{new33})).  But in this case the new $d=5$ operators
related to neutrino masses are likely to dominate and quite
naturally lead to lifetimes bounded above in the range of \((1-20)
\times 10^{33}\) years (as noted in Eq.~(\ref{new34})). \emph{Thus
in the presence of the new operators, the range of
\(10^{33}-10^{34}\) years for proton lifetime is not only very
plausible, but it also provides a reasonable upper limit for the
same, for the case of ESSM embedded in $SO(10)$.} \item We see
that the gauge boson mediated $d=6$ operators, for the case of
MSSM embedded in $SU(5)$ or $SO(10)$, though typically suppressed
compared to the $d=5$ operators, can lead to inverse decay rates
\(\Gamma^{-1}(p \rightarrow e^{+}\pi^{0})\) as short as about
$10^{34}$ years (see Eq.~(35)).  It should be stressed that the
$e^{+}\pi^{0}$-mode is the \emph{common denominator} of all GUT
models ($SU(5)$, $SO(10)$, etc.) which unify quarks and leptons
and the three gauge forces.  Its rate is determined essentially by
$\alpha_{H}$ and the SUSY unification scale, without the
uncertainty of the SUSY spectrum.  I should also mention that the
$e^{+}\pi^{0}$-mode is predicted to be the dominant mode in the
flipped \(SU(5) \times U(1)\)-model \cite{flip}, and also as it
turns out in certain higher dimensional GUT-models
\cite{higherdimGUT}, as well as in a model of compactification of 
M-theory on a manifold of $G_2$ holonomy
\cite{FriedmanWitten}. For these reasons, intensifying the search
for the $e^{+}\pi^{0}$-mode to the level of sensitivity of about
$10^{35}$ years in a next-generation proton decay detector should
be well worth the effort.

It may be noted that for the case of ESSM embedded in $SO(10)$ or
$SU(5)$, since $\alpha_{\mathrm{unif}}$ and the unification scale
(thereby the masses of the $X$, $Y$ gauge bosons) are raised by
nearly a factor of (6 to 7) and (2.5 to 5) respectively, compared
to those for MSSM (see Ref.~\cite{BabuJi} and discussions above),
while the inverse decay rate is proportional to
\((M_{X}^{4}/\alpha_{\mathrm{unif}}^{2})\) we expect
\[\Gamma^{-1}(p \rightarrow e^{+}\pi^{0})_{\mathrm{ESSM}}^{d=6}
\approx (\mbox{1 to 17}) \Gamma^{-1}(p \rightarrow
e^{+}\pi^{0})_{\mathrm{MSSM}}^{d=6}.\] This raises the interesting
possibility that for ESSM embedded in $SO(10)$, both
$\bar{\nu}K^{+}$ (arising from $d=5$) and $e^{+}\pi^{0}$ (arising
from $d=6$) can have comparable rates, with proton having a
lifetime $\sim (1/2 - 2) \times 10^{34}$ years.
\end{enumerate}

Before concluding I should mention that there have been several
old and new attempts in the literature based on compactification
of string/M-theory \cite{FriedmanWitten,seeCandelas}, as well as 
of a presumed 5D
or 6D point-particle GUT-theory \cite{SampleRefs1,SampleRefs2}, 
which project
out the color triplets (anti-triplets) belonging to
$\mathbf{5}_{H}$ ($\bar{\mathbf{5}}_{H}$) of $SU(5)$ or
$\mathbf{10}_{H}$ of $SO(10)$ from the massless spectrum in 4D,
through the process of compactification.  As a result, they obtain
a non-GUT SM-like gauge symmetry, and in some cases the $G(224)$
symmetry (see e.g. \cite{StringG(224)} and \cite{5DG(224)}) in 4D.
In the process they eliminate (often using discrete symmetries) or
strongly suppress the standard $d=5$ proton decay operators, though not
necessarily the $d=6$.

These approaches are interesting in their own right. There are, however, 
some constraints which should be satisfied if one wishes to understand 
certain observed features of low energy physics. In particular, it seems 
to me that at the very least
B-L should emerge as a gauge symmetry in 4D so as to protect the RH
neutrinos from getting a string-scale mass (see footnote~2) 
\emph{and} equally important to implement
baryogenesis via leptogenesis, as discussed in Secs.~3 and 4.  
This feature is not available in
models which start with $SU(5)$ in 5D, or in those that obtain only a standard
model-like gauge symmetry without B-L in 4D. Furthermore, the full
$SU(4)$ color symmetry, which of course contains B-L, plays a
crucial role in yielding not only (a) the (empirically favored)
relation \(m_{b}(M_{X}) \approx m_{\tau}\), but also (b) the relation
\(m(\nu_{\mathrm{Dirac}}^{\tau}) = m_{\mathrm{top}}(M_{X})\) which
is needed to account for the observed value of \(\Delta
m^{2}(\nu_{2}-\nu_{3})\) (see Sec.~3), and (c) the smallness of
$V_{cb}$ together with the near maximality of \(\sin^{2}
2\theta_{\nu_{\mu}\nu_{\tau}}^{\mathrm{osc}}\), as observed.  The
symmetry \(SU(2)_{L} \times SU(2)_{R}\) is also most useful in
that it relates the masses and mixings in the up and the down
sectors.  Without these correlations, the successful predictions
listed in Eq.~(16) will not emerge.

In short, as noted in Secs.~3 and 4, an understanding of neutrino
oscillations and leptogenesis as well as of certain intriguing
features of the masses and mixings of all fermions including
neutrinos seems to strongly suggest that minimally the symmetry
$G(224)$, or maximally the symmetry $SO(10)$, should survive as an
effective symmetry in 4D.  If the symmetry $G(224)$ rather than
$SO(10)$ survives in 4D near the string scale, the familiar color
triplets would be projected out through compactification [see e.g.
\cite{StringG(224)} and \cite{5DG(224)}].\footnote{The issue of
gauge coupling unification for this case is discussed in Sec.~2.}
In this case, there is no need for a doublet-triplet splitting
mechanism and the standard $d=5$ operators are either strongly
suppressed or completely eliminated. \emph{However, as long as the
Majorana masses of the RH neutrinos and the CKM mixings are
generated through the minimal Higgs system as in Sec.~3, the new
$d=5$ operators (Fig.~3) would still generically be present, and
would be the dominant source of proton decay.}  Like the standard
$d=5$ operators (Fig.~2), the new $d=5$ operators also lead to
$\bar{\nu}K^{+}$ and $\bar{\nu}\pi^{+}$ as dominant modes, but in
contrast to the standard operators, the new ones can lead to
prominent $\mu^{+}K^{0}$-modes \cite{BPW2} (see
Eq.~(\ref{new35})).

Given the empirical success of the supersymmetric
$G(224)/SO(10)$-framework, derivation of this framework, at least
that based on an effective $G(224)$-symmetry in 4D leading to the
pattern of Yukawa couplings as in Sec.~3, from an underlying
theory remains a challenge.  At the same time, based on its
empirical support so far, it makes sense to test this picture
thoroughly.  There are two notable pieces of this picture still
missing. One is supersymmetry, which will be tested at the LHC.
The other is proton decay.
\end{subsection}

\begin{subsection}{Section Summary}
Given the importance of proton decay, a systematic study of this
process has been carried out within a supersymmetric
$SO(10)/G(224)$ framework that successfully describes the fermion
masses, neutrino oscillations and leptogenesis.  Special attention
is paid in this study to the dependence of the d=5 proton decay
amplitude on the masses and mixings of the fermions and the
GUT-scale threshold effects.  Allowing for both the MSSM and the
ESSM alternatives within this $SO(10)/G(224)$ framework and
including the standard as well as the new $d=5$ operators, one
obtains (see Eqs.~(\ref{new31})--(\ref{new35})) a conservative upper limit on
proton lifetime given by:
\begin{equation}
\begin{array}{lr}
\tau_{\mathrm{proton}} \lsim (1/3 - 2) \times 10^{34} \mbox{ yrs}
& \left( \begin{array}{c} \mbox{SUSY} \\ SO(10)/G(224) \end{array} \right)
\end{array}
\label{new36}
\end{equation}
with $\bar{\nu}K^{+}$ and $\bar{\nu}\pi^{+}$ being the dominant modes and
quite possibly $\mu^{+}K^{0}$ being prominent.

The $e^+\pi^o$-mode induced by gauge boson-exchanges should have an 
inverse decay rate in the range of $10^{34}-10^{36}$ years (see Eq. 
(35)).  The implication of these predictions for a next-generation 
detector is emphasized in the next section.
\end{subsection}

\end{section}

\newpage
\begin{section}{Concluding Remarks}
In this talk, I have argued that but for two missing pieces -- supersymmetry 
and proton decay -- the evidence in favor of supersymmetric grand unification 
is now strong. It includes: (i) the observed family multiplet-structure, 
(ii) quantization of electric charge, (iii) the meeting of the three gauge 
couplings, (iv) neutrino oscillations (atmospheric and solar), (v) the 
intricate pattern of the masses and mixings of all fermions, including 
neutrinos, and (vi) the likely need for leptogenesis to account for the 
observed baryon asymmetry of the universe. All of these features can be 
understood simply and even quantitatively (see e.g.\ Eqs.~(3), (4), and 
(16)) within the concept of supersymmetric grand unification based on an 
effective string-unified $G(224)$ or $SO(10)$-symmetry in 4D. As discussed in 
Secs 3 and 4, attempts to understand especially (a) the tiny neutrino 
masses, (b) the baryon asymmetry of the universe (via leptogenesis), as 
well as (c) certain features of quark-lepton masses and mixings seem to 
select out the G(224)/SO(10) route to unification, as opposed to other 
alternatives.

A systematic study of proton decay has thus been carried out within this 
SO(10)/G(224) framework \cite{BPW1, JCPICTPtalk}, allowing for the 
possibilities of both MSSM and ESSM, and including the contributions for 
the gauge boson-mediated d=6, the standard d=5 as well as the new d=5 
operators related to the Majorana masses of the RH neutrinos. Based on 
this study, I have argued that a conservative upper limit on the lifetime 
of the proton is about $(\frac{1}{3}-2)\times 10^{34}$ years.

So, unless the fitting of all the pieces (i)-(vi) listed above is a mere 
coincidence, it is hard to believe that that is the case, discovery of 
proton decay should be around the corner. Allowing for the possibility 
that proton lifetime may well be near the upper limit stated above, a next 
generation detector, of the type proposed by UNO and Hyperkamiokande, 
providing a net gain in sensitivity by about a factor of five to ten, compared 
to SuperK, would thus be needed to produce real proton decay events and 
distinguish them from the background.

The reason for pleading for such improved searches is that proton
decay would provide us with a wealth of knowledge about physics at
truly short distances ($<10^{-30}$ cm), which cannot be gained by
any other means. Specifically, the observation of proton decay, at
a rate suggested above, with $\overline{\nu}K^{+}$ mode being
dominant, would not only reveal the underlying unity of quarks and
leptons but also the relevance of supersymmetry.  It would also
confirm a unification of the fundamental forces at a scale of
order $2\times10^{16}$ GeV. Furthermore, prominence of the
$\mu^{+}K^{0}$ mode, if seen, would have even deeper significance,
in that in addition to supporting the three features mentioned
above, it would also reveal the link between neutrino masses and
proton decay, as discussed in Sec. 5.  {\it In
this sense, the role of proton decay in probing into physics at
the most fundamental level is unique}. In view of how valuable
such a probe would be and the fact that the predicted upper limit
on the proton lifetime is at most a factor of three to ten higher
than the empirical lower limit, the argument in favor of building
an improved detector seems compelling. 

Such a detector should of course be designed to serve multiple goals 
including especially improved studies of neutrino oscillations and 
supernova signals. These ideas and others including that of a neutrino 
factory were discussed intensively at the NeSS meeting held recently in 
Washington \cite{NESS}.

To conclude, the discovery of proton decay would constitute a landmark in 
the history of physics. That of supersymmetry would do the same. The 
discoveries of these two features -- supersymmetry and proton decay -- 
would fill the two missing pieces of gauge unification and would 
shed light on how such a unification may be extended to include gravity 
in the context of a deeper theory. The question thus poses: Will our 
generation give itself a chance to realize \emph{both}?
\end{section}

{\bf Acknowledgments:} I would like to thank Gerard 't Hooft and Antonino 
Zichichi for the kind invitation to lecture at the Erice School. I have 
benefitted from many collaborative discussions with Kaladi S. Babu and Frank 
Wilczek on topics covered in this lecture. I am also grateful to Pasquale Di 
Bari and Qaisar Shafi for discussions on the topic covered in Sec. 4. The 
research presented here is supported in part by DOE grant no. 
DE-FG02-96ER-41015.

\bibliography{pati}

\begin{thebibliography}{100}
\bibitem{FN26}
For G(224), one can choose the corresponding sub-multiplets -- that is (1, 1,
  15)$_H$, (1, 2, $\bar{4}$)$_H$, (1, 2, 4)$_H$, (2, 2, 1)$_H$ -- together with
  a singlet $S$, and write a superpotential analogous to Eq. (\ref{eq:Yuk}).
\bibitem{newFN36}
Note that the magnitudes of $\eta$, $\epsilon$ and $\sigma$ are fixed by the
  input quark masses. Furthermore, one can argue that the two contributions for
  $\theta_{\nu_\mu\nu_\tau}^{\rm osc}$ [see Eq. (\ref{eq:pred})] necessarily
  add to each other as long as $|y|$ is hierarchical ($\sim 1/10$) \cite{BPW1}.
  As a result, once the sign of $\epsilon$ relative to $\eta$ and $\sigma$ is
  chosen to be negative, the actual magnitudes of $V_{cb}\approx (0.044)$ and
  $\sin^2 2\theta_{\nu_\mu\nu_\tau}^{\rm osc}\approx 0.92-0.99$ emerge as
  predictions of the model \cite{BPW1,JCPICTPtalk}.
\bibitem{FN32}
Note that such an operator would be allowed by the flavor symmetry defined in
  Ref.~\cite{FN9} if one sets $a=1/2$. In this case, operators such as $W_{23}$
  and $W_{33}$ that would contribute to $\nu_L^\mu\nu^\tau_L$ and
  $\nu_L^\tau\nu^\tau_L$ masses would be suppressed relative to $W_{12}$ by
  flavor symmetry.
\bibitem{FN38}
As an example, one such fit with complex parameters assigns
  \cite{BabuJCPCascais}: $\sigma=0.10-0.012\,i$, $\eta=0.12-0.05\,i$,
  $\epsilon=-0.095$, $\eta'=4.0\times 10^{-3}$, $\epsilon'=1.54\times 10^{-4}
  e^{i\pi/4}$, $\zeta_{22}^u=1.25\times 10^{-3}e^{i\pi/9}$ and
  $\zeta_{22}^d=4\times 10^{-3}e^{i\pi/2}$, ${\cal M}_u^0\approx 110$ GeV,
  ${\cal M}_D^0\approx 1.5$ GeV, $y\approx -1/17$ (compare with Eq.
  (\ref{eq:fit}) for which $\zeta_{22}^u=\zeta_{22}^d=0$). One obtains as
  outputs: $m_{b,s,d}\approx(\mbox{5 GeV, 132 MeV, 8 MeV})$, $m_{c,u}\approx
  (\mbox{1.2 GeV, 4.9 MeV})$, $m_{\mu, e}\approx(\mbox{102 MeV, 0.4 MeV})$ with
  $m_{t,\tau}\approx(\mbox{167 GeV, 1.777 GeV})$, $(V_{us}, V_{cb}, |V_{ub}|,
  |V_{td}|)\approx(0.217, 0.044, 0.0029, 0.011)$, while preserving the
  predictions for neutrino masses and oscillations as in Eq. (\ref{eq:pred}).
  The above serves to demonstrate that complexification of parameters of the
  sort presented above can preserve the successes of Eq. (\ref{eq:pred})
  (\cite{BPW1}). This particular case leads to $\eta_W=0.29$ and
  $\rho_W=-0.187$ \cite{BabuJCPCascais}, to be compared with the corresponding
  standard model values (obtained from $\epsilon_K$, $V_{ub}$ and $\Delta
  m_{Bd}$) of $(\eta_W)_{\rm SM}\approx 0.33$ and $(\rho_W)_{\rm SM} \approx
  +0.2$. The consistency of such values for $\eta_W$ and $\rho_W$ (especially
  reversal of the sign of $\rho_W$ compared to the SM value), in the light of
  having both standard model and SUSY-contributions to CP and
  flavor-violations, and their distinguishing tests, are discussed in
  Ref.~\cite{BabuJCPCascais}.
\bibitem{newFN50}
The factor 0.7 in Eq. (\ref{eq:d}) [instead of 1 in Eq. (13) of the first paper
  in Ref.~\cite{Bari}] is an estimate that incorporates the modification needed
  for SUSY corresponding to a doubling of $N_1$-decay width owing to the
  presence of both $l+H$ and $\tilde l+\tilde H$-modes and an increase of $g^*$
  from 106 for the standard model to 228 for SUSY.
\bibitem{newFN51}
One can verify that $K\equiv (\Gamma(N_1)/2H)_{T=M_1} \approx (0.37)[M_{\rm
  Pl}/(1.66 \sqrt{g^*}(8\pi v^2))]\tilde{m_1}\approx 234 (\tilde{m_1}/{\rm
  eV})$, where 0.37 denotes the usual time-dilation factor, $g^*{\rm
  (for~SUSY)}\approx 230$ and $v\approx 174$ GeV. For comparison, we note that
  if one includes only inverse decays (thus neglecting $\Delta L\neq
  0$-scatterings) in the Boltzmann equations, one would obtain: $\kappa\approx
  0.3/[K(\ln K)^{0.6}]$ for $K>10$ \cite{d}, and $\kappa\approx 1/2K$ for
  $1\lsim K\lsim 10$. As pointed out in Ref.~\cite{Bari}, these expressions,
  frequently used in the literature, however, tend to overestimate $\kappa$ by
  nearly a factor of 7. In what follows, we will therefore use Eq. (\ref{eq:d})
  to evaluate $\kappa$.
\bibitem{M2Vevs}
Incorporating such an effective superpotential in accord with the assignment of
  flavor-charges suggested in Refs. \cite{FN9} and \cite{FN33} would involve
  two additional singlets with appropriate charges. The (VEV)$^2$ of one or
  both of these may represent $M^2$. Derivation of such a picture with
  appropriate flavor-charge assignments from an underlying (string/M) theory
  still remains a challenge.
\bibitem{NonThermalRelaxedGravitino}
Note that for this non-thermal case, since the gravitino-constraint is relaxed,
  $N_1$ can be chosen heavier than for the case considered before (the thermal
  case), still in accord with Eq. \eqref{eq:MajM}. Since $Y_B\propto\epsilon_1
  T_{\rm RH}/{m_{\rm infl}}$, while $\epsilon_1\propto (M_1/M_2)$, $T_{\rm
  RH}\propto M_1 ({m_{\rm infl}})^{1/2}$ and ${m_{\rm infl}}\propto\lambda$, we
  see that $Y_B\propto (M_1^2/M_2)/\sqrt{\lambda}$, for a constant $M$, for the
  case of non-thermal leptogenesis.
\end{thebibliography}
\bibliographystyle{unsrt}

\newpage

\begin{figure}[t]
\begin{center}
\begin{picture}(200,100)(0,0)
 \ArrowLine(20,10)(60,50)
  \ArrowLine(20,90)(60,50)
   \ArrowLine(180,10)(140,50)
    \ArrowLine(180,90)(140,50)
\Gluon(60,50)(140,50){6}{7} \Text(30,10)[c]{$q$}
\Text(30,90)[c]{$q$} \Text(170,90)[c]{$q$} \Text(170,10)[c]{$l$}
\Text(100,63)[c]{$X,Y$}
\end{picture}
\caption{\label{Fig1}$d=6$ proton decay operator.}
\end{center}
\end{figure}
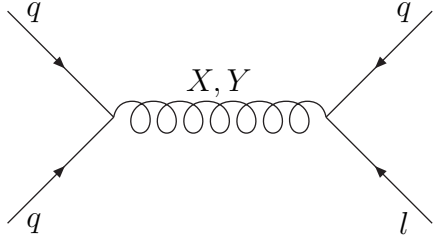

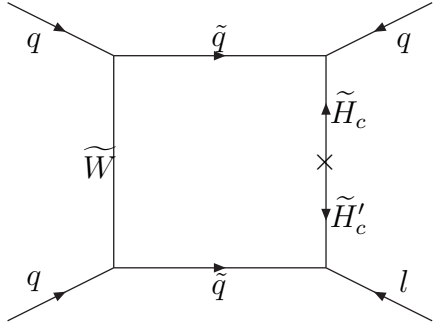
\begin{figure}[t]
\begin{center}
\begin{picture}(200,140)(0,0)
 \ArrowLine(20,10)(60,30)
  \ArrowLine(20,130)(60,110)
   \ArrowLine(180,130)(140,110)
    \ArrowLine(180,10)(140,30)
\Text(30,25)[c]{$q$} \Text(30,115)[c]{$q$} \Text(170,115)[c]{$q$}
\Text(170,25)[c]{$l$}

 \ArrowLine(60,110)(140,110)
  \Line(60,30)(60,110)
  \ArrowLine(60,30)(140,30)
   \ArrowLine(140,70)(140,110)
    \ArrowLine(140,70)(140,30)
    \Text(140,70)[c]{$\times$}

\Text(55,70)[c]{$\widetilde{W}$} \Text(100,117)[c]{$\tilde{q}$}
\Text(100,23)[c]{$\tilde{q}$}
\Text(149,50)[c]{$\widetilde{H}'_c$}\Text(149,90)[c]{$\widetilde{H}_c$}
\end{picture}
\caption{\label{Fig2}The standard $d=5$ proton decay operator. The
$\widetilde{H}'_c$ ($\widetilde{H}_c$) are color
triplet(anti-triplet) Higgsinos belonging to
$5_H$($\overline{5}_H$) of $SU(5)$ or $10_H$ of  $SO(10)$.}
\end{center}
\end{figure}

\begin{figure}[t]
\begin{center}
\begin{picture}(200,140)(0,0)
 \ArrowLine(20,10)(60,30)
  \ArrowLine(20,130)(60,110)
   \ArrowLine(180,130)(140,110)
    \ArrowLine(180,10)(140,30)
\Text(30,25)[c]{$q$} \Text(30,115)[c]{$q$} \Text(170,115)[c]{$q$}
\Text(170,25)[c]{$l$}
 \ArrowLine(60,110)(140,110)
  \Line(60,30)(60,110)
  \ArrowLine(60,30)(140,30)
   \ArrowLine(140,70)(140,110)
    \ArrowLine(140,70)(140,30)
    \Text(140,70)[c]{$\times$}
 \Text(55,70)[c]{$\widetilde{W}$}
\Text(100,117)[c]{$\tilde{q}$} \Text(100,23)[c]{$\tilde{q}$}
\Text(151,50)[c]{$16_H$}\Text(151,90)[c]{$\overline{16}_H$}
\DashArrowLine(100,80)(140,110){2}
 \DashArrowLine(100,60)(140,30){2}
\GCirc(100,80){3}{0.7} \GCirc(100,60){3}{0.7}
\Text(110,99)[c]{$\overline{16}_H$}
 \Text(110,43)[c]{$16_H$}
\end{picture}
\caption{\label{Fig3}The ``new'' $d=5$ operators related to the
Majorana masses of the RH neutrinos. Note that the vertex at the
upper right utilizes the coupling in Eq.(9) which assigns Majorana
masses to $\nu_R$'s, while the lower right vertex utilizes the
$g_{ij}$ couplings in Eq.(8) which are needed to generate CKM
mixings.}
\end{center}
\end{figure}
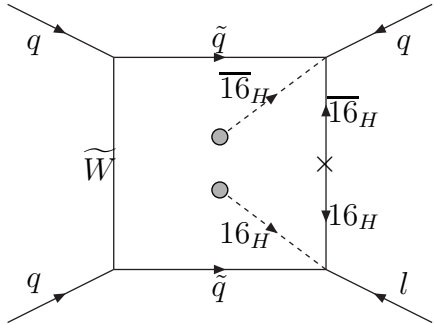

\end{document}